\documentclass[acmsmall,screen]{acmart} 
\AtBeginDocument{%
  \providecommand\BibTeX{{%
    \normalfont B\kern-0.5em{\scshape i\kern-0.25em b}\kern-0.8em\TeX}}}

\setcopyright{acmcopyright}
\copyrightyear{2021}
\acmYear{2021}
\acmDOI{10.1145/1122445.1122456}

\acmJournal{CSUR}
\acmVolume{37}
\acmNumber{4}
\acmArticle{111}
\acmMonth{12}

\usepackage{caption}
\usepackage{subcaption}
\begin{document}

\title{When Creators Meet the Metaverse: A Survey on Computational Arts}



\author{Lik-Hang Lee}\authornote{\textbf{The Corresponding Author}: Lik-Hang Lee, \url{lhleeac@connect.ust.hk}. \textbf{Co-second Authorship}: Zijun Lin, Rui Hu, Zhengya Gong, Abhishek Kumar and Tangyao Li contribute equally in this article. 
\textbf{Second Affiliations}: Pan Hui is also affiliated with Department of Computer Science, University of Helsinki, Finland. Zijun Lin, Tangyao Li, Sijia Li conducted this research with Augmented Reality and Media Lab at KAIST.
\textbf{Acknowledgements}: This research has been supported in part by 5GEAR (Decision No. 318927) and FIT (Decision No. 325570) projects funded by the Academy of Finland, the Nokia Foundation, and China Scholarship Council (No. 202107960006). We also thank you for Ms. Zhihan Wang's article sharing.} 
\affiliation{%
  \institution{Augmented Reality and Media Lab, KAIST}
  \country{South Korea}
}

\author{Zijun Lin}
\affiliation{%
  \institution{Department of Economics, University College London}
  \country{United Kingdom}
}

\author{Rui Hu}
\affiliation{%
  \institution{Computational Media \& Arts, The Hong Kong University of Science and Technology}
  \country{Hong Kong}
}

\author{Zhengya Gong}
\affiliation{%
  \institution{Center for Ubiquitous Computing, University of Oulu}
  \country{Finland}
}

\author{Abhishek Kumar}
\affiliation{%
  \institution{Department of Computer Science, University of Helsinki}
  \country{Finland}
}

\author{Tangyao Li}
\affiliation{
  \institution{School of Mathematical Sciences, Queen Mary University of London}
  \country{United Kingdom}
}

\author{Sijia Li}
\affiliation{
  \institution{Department of Architecture, Inha University}
  \country{South Korea}
}


\author{Pan Hui}
\affiliation{%
   \institution{Computational Media \& Arts, The Hong Kong University of Science and Technology}
  \country{Hong Kong}
}






\renewcommand{\shortauthors}{Lee, et al.}

\begin{abstract}

The metaverse, enormous virtual-physical cyberspace, has brought unprecedented opportunities for artists to blend every corner of our physical surroundings with digital creativity. This article conducts a comprehensive survey on computational arts, in which seven critical topics are relevant to the metaverse, describing novel artworks in blended virtual-physical realities. 
The topics first cover the building elements for the metaverse, e.g., virtual scenes and characters, auditory, textual elements. Next, several remarkable types of novel creations in the expanded horizons of metaverse cyberspace have been reflected, such as immersive arts, robotic arts, and other user-centric approaches fuelling contemporary creative outputs.
Finally, we propose several research agendas: democratising computational arts, digital privacy and safety for metaverse artists, ownership recognition for digital artworks, technological challenges, and so on.
The survey also serves as introductory material for artists and metaverse technologists to begin creations in the realm of surrealistic cyberspace.
\end{abstract}



\makeatletter
\let\@authorsaddresses\@empty
\makeatother

\maketitle

\section{Introduction}
\label{sec:introduction}

Art is defined as a vastly diversified range of human activities that create products of imagination and creativity in various channels, including but not limited to visual, auditory, dancing, theatrical performance, poetry, artefacts and sculpture, and other physical objects~\cite{sep-art-definition}. 
In the broadest sense, artists, or equivalently creators, who engaged in the artistic creation process, leverage various materials, techniques, and forms to express their ideas and observations and communicate their feelings and thoughts with their audiences~\cite{Abell2012ArtWI}. 
Since the popularisation of computers in the 1970s, artists and other digital pioneers have made many trials of utilising computational devices to express their ideas and thoughts. To the best of our knowledge, the earliest publicly shown pieces of computational arts can be found in 
\textit{Georg Nees: Computergrafik}, 
exhibited in February 1965 in Stuttgart, Germany\footnote{\url{https://www.historyofinformation.com/detail.php?entryid=3921}}. 
From then on, computers have energised an available set of alternatives and instruments and hence inventively the landscape of creative processes. Specifically, various computer-mediated environments, primarily supported by digitalised media and technology gadgets, have enriched the creative process. As a result, the new process generates enormously novel yet interactive artworks that have placed themselves under the larger umbrella term ‘Computational Media and Arts’.

\begin{figure}[!h]
     \centering
     \begin{subfigure}[b]{0.3\textwidth}
     \centering
	\includegraphics[width=\linewidth]{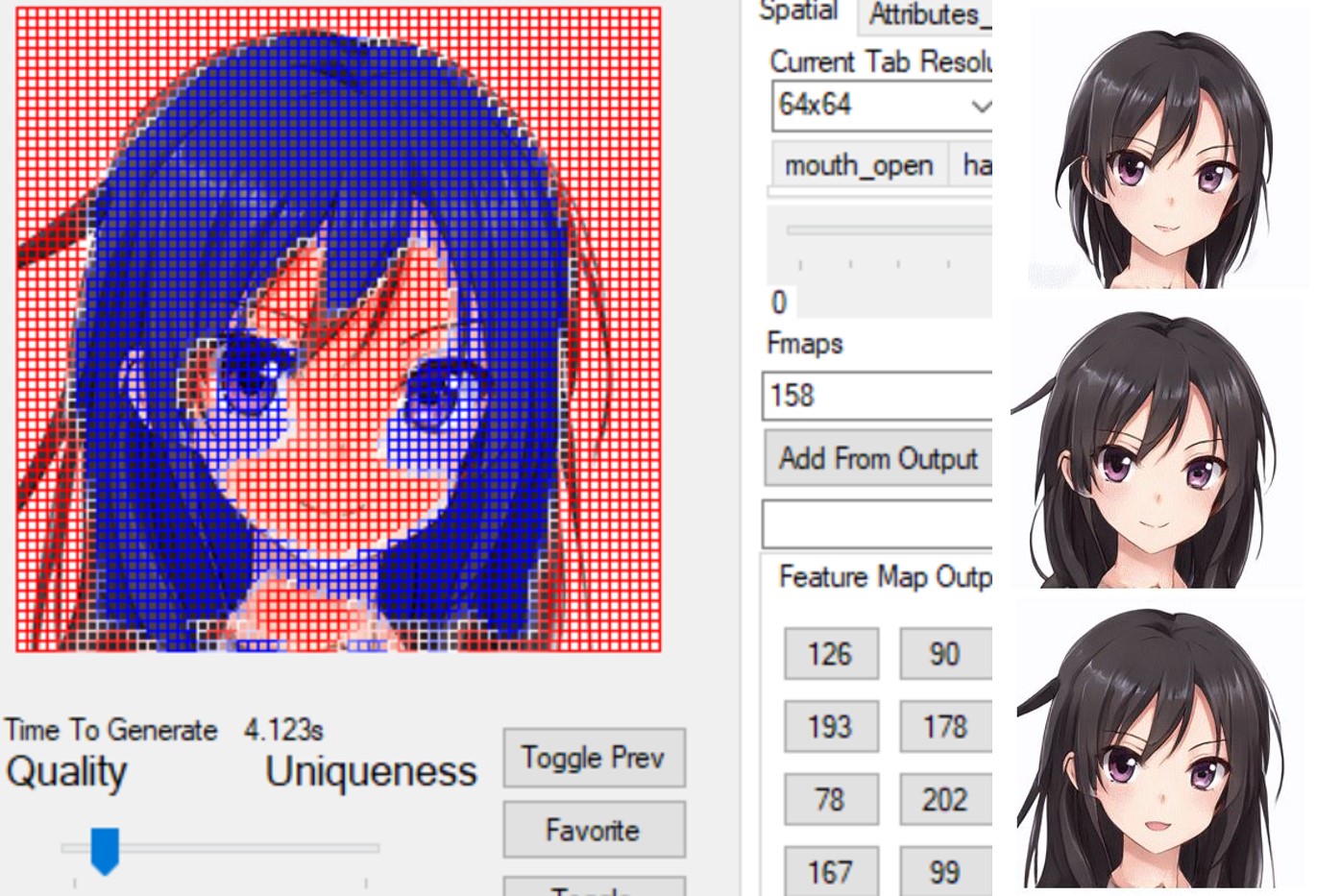}
	\caption{Creating a new anime character by adjusting the scroll bars to change its hairs and mouths.}
	\label{fig:gan-anime}
     \end{subfigure}
     \hfill
     \begin{subfigure}[b]{0.37\textwidth}
	\centering
	\includegraphics[width=\linewidth]{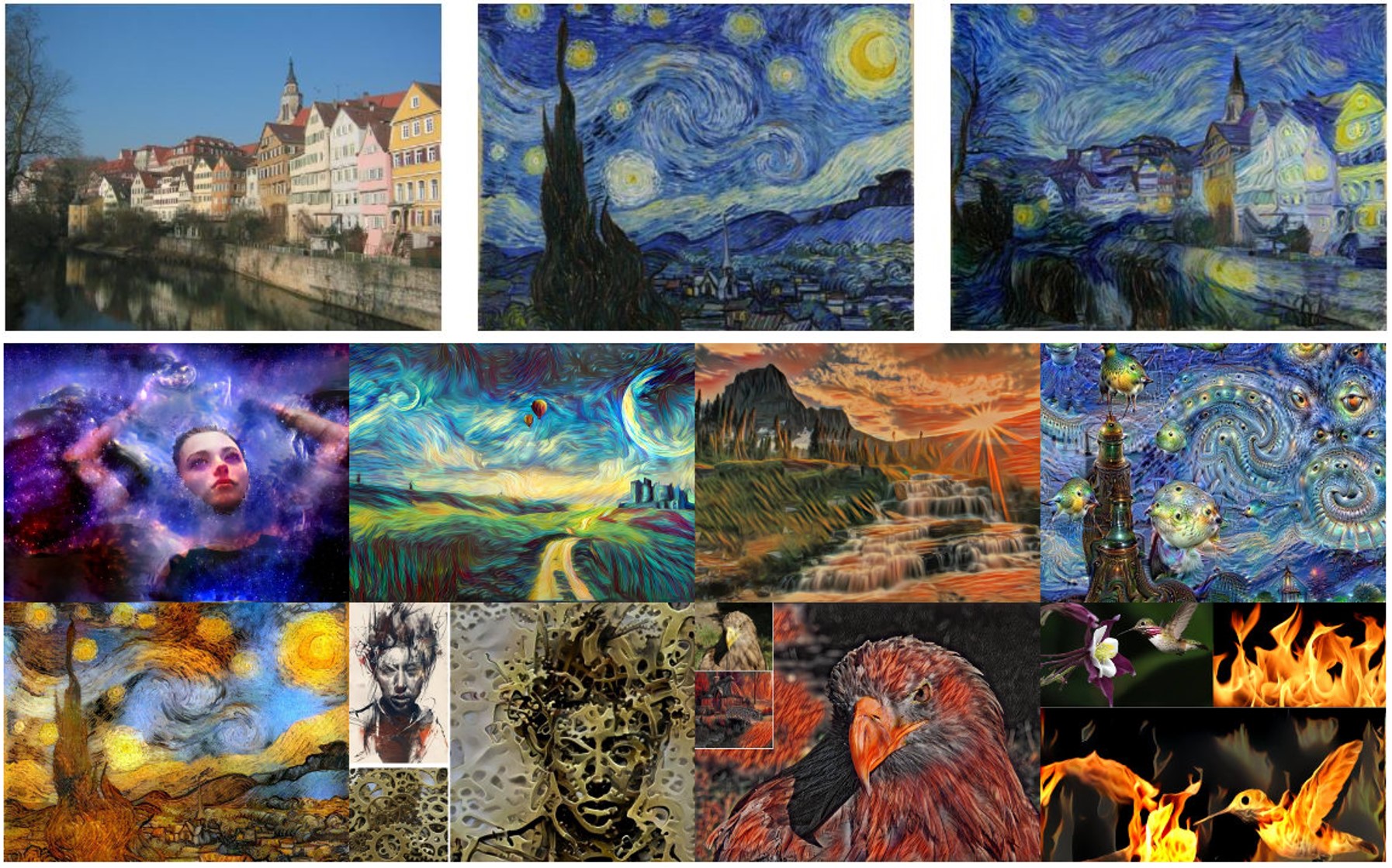}
	\caption{Style transfer enables computer-generated paintings.}
	\label{fig:gan-paint}
     \end{subfigure}
     \hfill
     \begin{subfigure}[b]{0.3\textwidth}
	\centering
	\includegraphics[width=.93\linewidth]{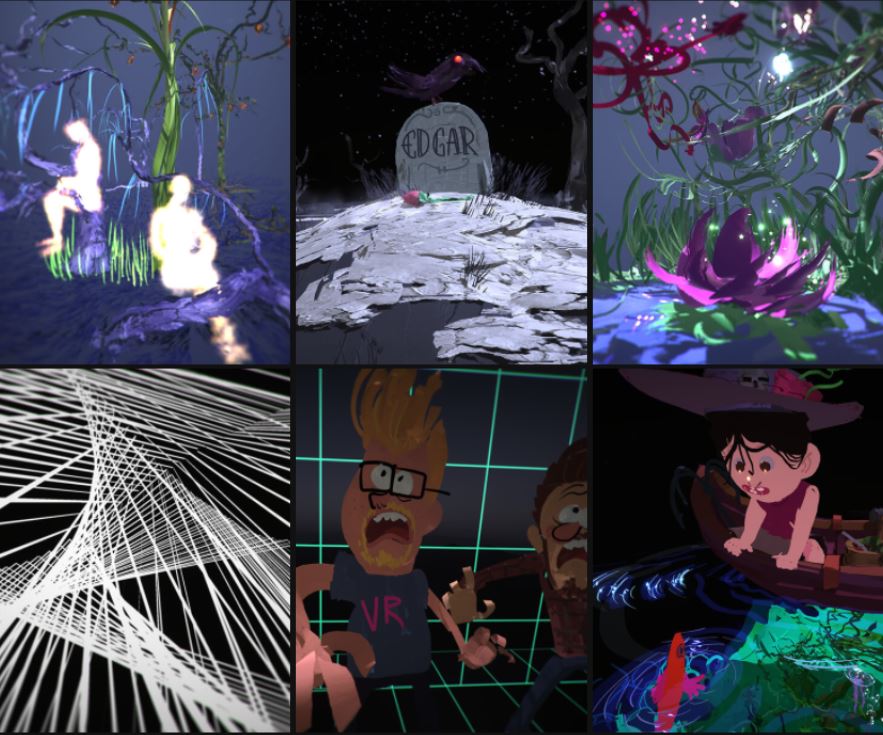}
	\caption{Creation of 3D artworks in virtual reality named Tilt Brush.}
	\label{fig:vr-paint}
     \end{subfigure}
        \caption{Recent examples of computational arts}
        \label{fig:intro}
\end{figure}

After the debut in the mid-20$^{th}$ Century, computational arts have travelled a long way. 
The role of computational arts is no longer limited to the digitisation of artworks and cost-effective enablers of pixelised artworks (e.g., creating animation on a drawing tablet). 
Nowadays, computers can actively participate in the creation process with the creators. 
Remarkably, the advancement of artificial intelligence (AI) also drastically changed the way we create artwork. 
For instance, an AI-enabled tool\footnote{\url{https://github.com/nolan-dev/GANInterface}}, supported by a StyleGAN model, allows the creations of new anime characters by intuitively adjusting the parameters of mouth, eyes, and hair (Figure~\ref{fig:gan-anime}).
Similarly, another well-known example, Deep Art\footnote{\url{https://deepart.io}}, is turning photography into a painting through extracting and learning the artistic styles of existing artworks (Figure~\ref{fig:gan-paint}).
It is important to note that AI is not the sole facilitator of computational arts, albeit we spot new opportunities for human-AI collaboration for artistic creation. 
We have witnessed other new forms of artistic representations. They have appeared as buildings, mechanics, robotics and drones, as well as virtual and augmented realities. 



In 2020, the COVID-19 pandemic significantly impacted our lifestyle. The digital transformation has been expedited by years, and its impact can be long-haul~\cite{continual-remote} -- people study and work remotely and spend their leisure time with digital tools and virtual platforms~\cite{10.1145/3411764.3445428}. As such, opportunities emerge for computational arts in the aftermath of this global crisis. 
While the COVID pandemic has engendered a sizeable shock to industries like filmmaking, where the employment in motion picture and sound recording in the US has dropped by more than 40\% in 2021\cite{navarro_2021}, other industries have remained robust. Notably, in the midst of the pandemic, the gaming company Nintendo's total revenue for 2021 is 46.5\% higher than in 2019 \cite{nintendo_market}. On an aggregate level, the video game industry has generated total revenue of 155.89 billion USD in 2020 and is projected to reach 268.81 billion USD by 2025 \cite{game_market}. 
In incipient fields of arts technologies, there are also signs of a promising future. In digital arts, more creators and traders are paying attention to the rise of the Non-Fungible Tokens (NFT) market, where its sales volume reached 2.5 billion USD two quarters into the year 2021 \cite{howcroft_2021}. The expansion of this new market not only supports digital content creators to claim rewards for their efforts, but the rising popularity of digital art also enables a broader range of participation from our society. In addition, traditional auctioning platforms like Sotheby's\footnote{\url{https://bit.ly/3B5i3KB}} has also embraced the new trend by holding online auctions for NFT artworks (More details 
in Section~\ref{sec:virtual art trading}).

Moreover, the increasing accessibility following the commercialisation of Augmented and Virtual Reality allows more to experience arts technologies. For instance, Tilt Brush\footnote{\url{https://www.tiltbrush.com/}} launched by Google, allows users to produce creative artworks with VR headsets (Figure~\ref{fig:vr-paint}). The relationship between new arts technologies and traditional arts is not in dichotomy. Instead, they can co-exist in harmony. Unlike some critics may posit, experts and the public's perception on what counts as an excellent digital artwork may coincide \cite{10.1145/3402443}. Moreover, the rise of NFTs can be interpreted as widening the aggregate art community as it encourages participation from the young generation~\cite{franceschet2021crypto}. 


The COVID-19 pandemic also confirms the people's acceptance to live and play with higher involvement of cyberspaces and virtual spaces. Therefore, Apple, Meta (Facebook, before re-branding) and Microsoft have launched their strategic plans to enter the era of the metaverse\footnote{\url{https://medium.com/building-the-metaverse/clash-of-the-metaverse-titans-microsoft-meta-and-apple-ce505b010376}}. 
In brief, the metaverse refers to gigantic and open 3D virtual spaces that allow an unlimited number of users to socialise, learn, work, collaborate, create and play in such cyberspaces~\cite{Lee2021AllON}. The concept of the metaverse has received tremendous attention from all walks of life. It is projected that the metaverse requires various new creation to build the 3D virtual worlds and user interactions. As stated in~\cite{Lee2021AllON}, content creations is a critical ecosystem factor to the sustainability of the metaverse. The metaverse users create new content in virtual spaces. That is, every participant in the metaverse, instead of professional contributors, would become a digital creator, and accordingly, an computational artist. 

With such projected accelerated digital transformation, namely `\textit{Digital Big Bang}' with the metaverse~\cite{Lee2021AllON}, the world merges seamlessly with virtual entities composed of various artistic components, for instance, 2D images, 3D objects, auditory effects, and so on. 
Remarkably, the metaverse demonstrates broader cyberspace extended to tangible and intelligent objects, such as robotics and flying drones. Thus, right now, there is a massive opportunity for digital creators or computational artists to explore novel ways of creating arts with computer-mediated environments, technology gadgets, robots and drones, and to name but a few. In the coming paragraphs, we briefly discuss several examples to illustrate the relationship between artworks and the metaverse.

\subsection{When Creators meet The Metaverse}

\begin{figure}[!h]
     \centering
     \begin{subfigure}[b]{0.29\textwidth}
     	\centering
	\includegraphics[width=.72\linewidth]{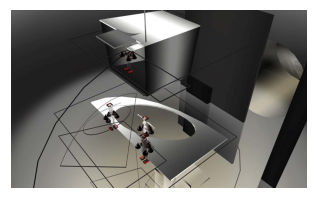}
	\caption{EI Lissitzky's (re)constructed Proun \#5A in the metaverse.} 
	\label{fig:Proun}
     \end{subfigure}
     \hfill
     \begin{subfigure}[b]{0.37\textwidth}
		\centering
	\includegraphics[width=\linewidth]{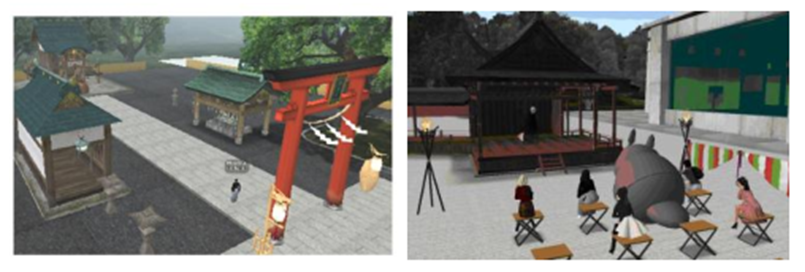}
	\caption{Constructing situated-learning metaverse platform for Japanese culture.}
	\label{fig:japanese}
     \end{subfigure}
     \hfill
     \begin{subfigure}[b]{0.31\textwidth}
	\centering
	\includegraphics[width=.78\linewidth]{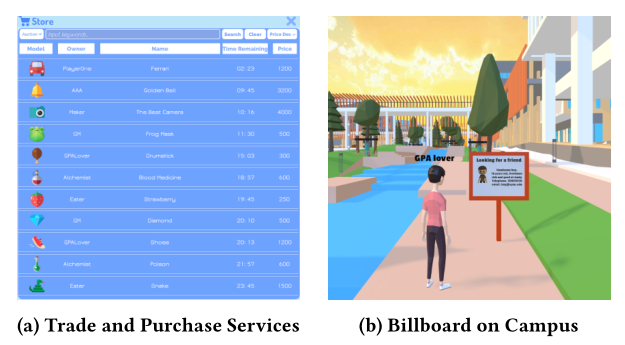}
	\caption{Metaverse ecosystems \& trading in the CUHK campus.}
	\label{fig:cuhksz-metaverse}
     \end{subfigure}
        \caption{Virtual worlds in the metaverse.}
        \label{fig:metaverse-intro}
\end{figure}

Earlier works tried to employ 3D virtual space to conceptualise creative artworks (e.g., visualisation of mysterious ideas~\cite{7398423},  communicate abstract concepts of biological genomics~\cite{Calvio2019ComputationalAI}.~\cite{Calvio2019ComputationalAI}), and test novel ideas (e.g., adding abstract artworks in architectures). 
In~\cite{7398423}, the magical nature named `\textit{the Aleph}' were represented as a cyberspace being, in which a number of Cartesian typographic elements `\textit{Azimuth}' attempts to present the unfamiliar concept in easy-to-understand yet artistic manners. In addition, computational artists work with scientists to visualise genome sequence data in aesthetic and comprehensible ways to the general public~\cite{Calvio2019ComputationalAI}. 
Moreover, the El Lissitzky's architecture in 1919, namely ``Proun \#5A''(Figure \ref{fig:Proun}), was recreated in a 3D virtual environment that allows 3D-object rendering and repositioning of multiple paintings or drawings~\cite{6980790}. 





The latest technology further pushes the boundary towards the realisation of the metaverse that pinpoints the key concept of digital twins and ultimately virtual-physical
105
 co-existence~\cite{Lee2021AllON} for multi-user socialisation and co-creation.  Through the lens of augmented reality (AR) headsets such as Meta Version 2~\cite{10.1145/2933540.2934351}, users can see virtual-physical blended worlds, in which virtual content superimposing on the top of physical counterparts of smart cities~\cite{csur-lee2022}. On the other hand, users with virtual reality (VR) headsets can situate in some immersive environments. As such, users, as represented by their avatars, will interact with other avatars in virtual cyberspaces inside the metaverse, i.e., the Second Life~\cite{5474848}. Such interaction can involve item enquiry and selection~\cite{10.1145/1643928.1643976}, and hence numerous items should first be created by multiple metaverse creators. 

Between the two ends of the spectrum (AR and VR), mixed reality (MR) requires high levels of scene understanding. This enables virtual objects (including virtual art) to seamlessly interact with the counterpart of physical objects. 
Decentralised governance of scene information requires an effective and real-time information sharing scheme among peer creators in a virtual-physical shared space~\cite{10.1145/3174910.3174952}. Meanwhile, sensors in real life can get information from the physical world and hence impacts virtual projection in virtual worlds~\cite{6093659}. Based on the above infrastructure, creators in the metaverse can socialise with each other in the virtual worlds, and achieve their common goals in situational contexts~\cite{10.1145/1952712.1952717}, e.g., creating new artistic contents that serve the goal of expressing Japanese culture and architectures (Figure~\ref{fig:japanese})~\cite{2011Constructing}. Additionally, the metaverse is characterised by a virtual marketplace, driven by blockchain technology, which facilitates peer-to-peer item trading (Figure~\ref{fig:cuhksz-metaverse}) in the metaverse~\cite{10.1145/3474085.3479238}. 
It is worthwhile to mention that the trading of (artistic) items is the fundamentals of building a metaverse community, in which creators or artists will spend a significant amount of time to create novel and creative contents in the metaverse.
Further discussion of virtual artwork trading is available in the next paragraph (Section~\ref{sec:virtual art trading}).


\subsection{An Artistic Metaverse in Embryo: Trading of Virtual Arts}
\label{sec:virtual art trading}
With the ease of duplicating and disseminating digital and media arts, many argue the orthodoxy valuation of artwork (i.e., based on scarcity and exclusiveness) is less relevant for this new type of creative work \cite{burns2001economic, burns2010valuation}. However, Non-Fungible Tokens shed light on the conundrum associated with digital artworks' inherent lack of scarcity and authentication \cite{mcconaghy2017visibility}. At the heart of virtual art trading, there lies the Non-Fungible Token (NFT), an eminent concept where its appeals go beyond the CMA community. Powered by blockchain technology, NFT can establish publicly recognisable ownership in virtual objects that are susceptible to piracy and forgery. The means by which such ownership is enshrined should not be unfamiliar to readers with some knowledge in cryptocurrency. Records of ownership in NFT, congruous to cryptocurrency, are logged by countless devices across the globe in a decentralised manner. 
In simpler terms, instead of storing the actual artwork (e.g., an image of a popular meme) like backing up data on a server, blockchains log records of `who\footnote{It is worth noting that the word `who' should not be taken explicitly. Instead, it is individuals' wallets addresses that get recorded.} owns what' pointing towards certain objects, where individuals cannot freely subvert these records. Therefore, artists who work on physical arts (such as large size art installations) may also mint new tokens to serve as certificates of ownership, where they can then offer these tokens to art collectors.  A close analogy can be where companies including Verisart and Artory take the initiative to utilise the blockchain to verify physical artworks \cite{whitaker2019art}. Though this idea may seem radical at first glance, it does possess some intrinsic appeals towards art gallery and private collectors who wish to have their collection displayed publicly or simply lacks the storage capacity. Additionally, it may also benefit art traders who expect high turnover for their physical artworks as it saves them the transportation cost of shipping the works back and forth. In return, some intermediary may store the physical artworks with a reasonable charge for the traders who frequently exchange ownership through NFTs. Analogously, nascent physical art forms like robotic art may also benefit from a wider acceptance of this new business paradigm. 


\begin{figure}[!h]
     \centering
     \begin{subfigure}[b]{0.49\textwidth}
        	\centering
	\includegraphics[width=.9\linewidth]{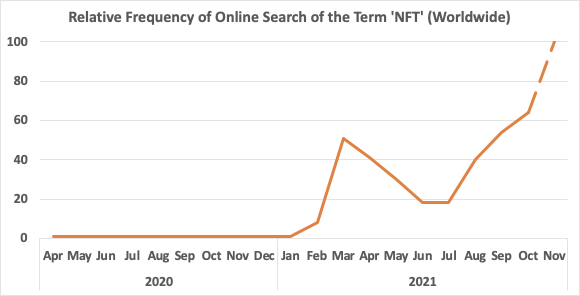}
	\caption{Rising Popularity of NFTs (Scale: 0--100)\protect\footnotemark.}
	\label{fig:NFT_frequency}
     \end{subfigure}
     \hfill
     \begin{subfigure}[b]{0.49\textwidth}
	\centering
	\includegraphics[width=.867\linewidth]{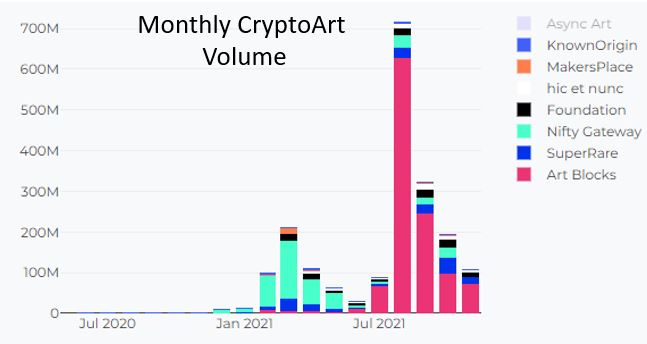}
	\caption{Crypto Arts trading activities emerge since 2021\protect\footnotemark.}
	\label{fig:v-ss}
     \end{subfigure}
        \caption{(a) Changes in the worldwide relative search frequency of `NFT' on Google and Youtube. 
        The dashed line indicates only partial information is available. (b) The trading of Crypto Arts via main platforms.}
        \label{fig:transfer-tras2}
\end{figure}
\addtocounter{footnote}{-1}
\footnotetext{{\url{https://bit.ly/3qsyijx}}}
\addtocounter{footnote}{1}
\footnotetext{\url{https://cryptoart.io/data}}

\begin{figure}[!h]
     \centering
    \begin{subfigure}[b]{0.3\textwidth}
     \centering
     \includegraphics[width=.8\linewidth]{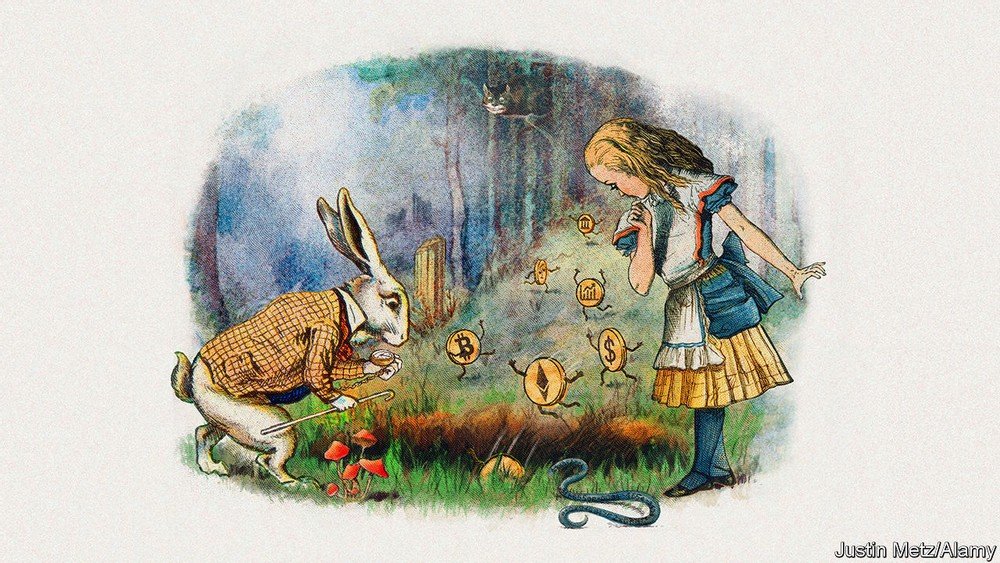}
	\caption{The Economist's front page image of `Alice in Wonderland' sold 
	as an NFT\protect\footnotemark.}
	\label{fig:alice}
     \end{subfigure}
     \hfill
     \begin{subfigure}[b]{0.3\textwidth}
     \centering
     \includegraphics[width=.8\linewidth]{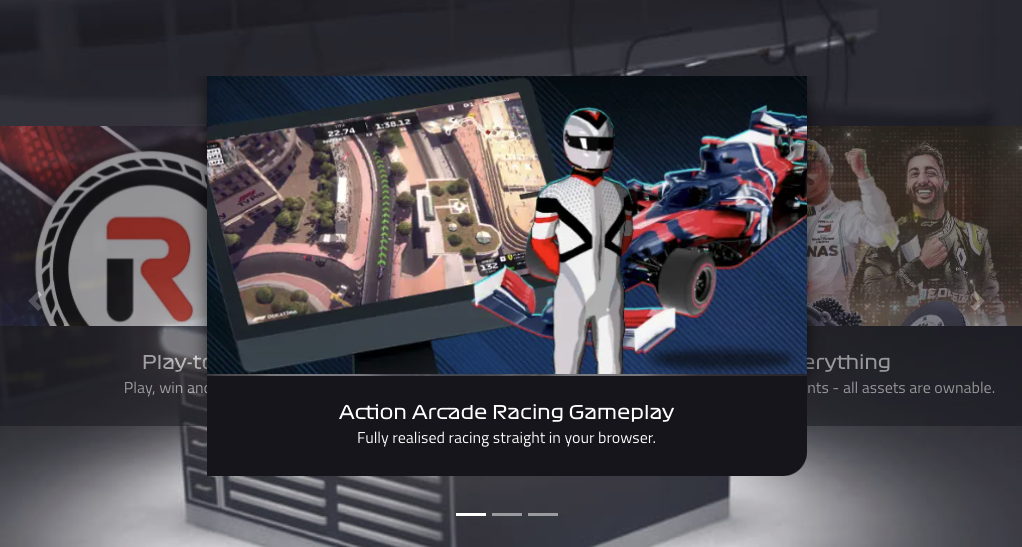}
	\caption{F1\textsuperscript{®} Delta Time: players collect and exchange in-game elements as NFTs\protect\footnotemark.} 
	
	\label{fig:F1}
     \end{subfigure}
     \hfill
     \begin{subfigure}[b]{0.3\textwidth}
		\centering
	\centering
	\includegraphics[width=.6\linewidth]{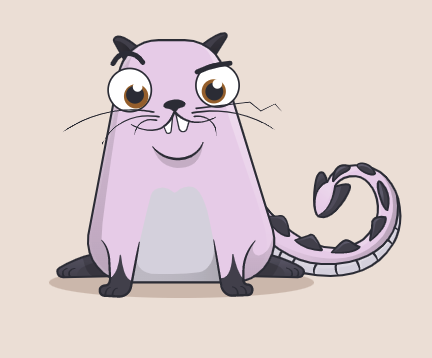}
	\caption{The cryptokitty that was sold for 600 ETH\protect\footnotemark.}
	\label{fig:cryptokitties}
     \end{subfigure}
        \caption{Virtual Arts in the metaverse.}
        \label{fig:virtual-trading}
\end{figure}

As a sharp contrast to a significant proportion of art dealers' conservative attitudes towards new technologies, say, virtual galleries through VR devices \cite{artbasel}, the market has been more willing to embrace NFTs in recent years. In complementary to the growing public interest in Non-Fungible Tokens as we demonstrate in Figure~\ref{fig:NFT_frequency} and rising NFT user count reflected by the amount of user wallets~\footnote{\url{https://bit.ly/31Ybn5k}}, developed marketplaces for digital artworks powered by NFTs already exist. On these platforms, creators offer a broad-spectrum of objects~\cite{Lee2021AllON}, including but not limited to drawings, avatars, audio and 3D models
. OpenSea, as an example of one of the leading NFT trading platforms, is valued at 1.5 billion USD \cite{matney_2021}, with exponential growth in trade volume since its birth\footnote{\url{https://bit.ly/3BU29mR}}. Successful adoption in NFT-based digital arts 
takes place outside the common NFT art marketplaces: the Cryptokitties, one of the first NFT collectable projects, received universal attention with one rare digital cat being sold at a jaw-dropping price of 600 ETH. Figure~\ref{fig:cryptokitties} shows the cryptokitty that was sold for 600 ETH~\cite{evans2019cryptokitties}. Likewise, the blockchain-based game F1\textsuperscript{®} Delta Time, illustrated in Figure \ref{fig:F1}, also achieved a similar triumph by releasing a limited-edition virtual race car. Some refer to this class of NFTs commonly seen in blockchain-based games as `smart collectables', which are digital tokens that enable additional functionality beyond plain ownership \cite{fai2021smart}. Once again, using the game F1\textsuperscript{®} Delta Time as an example, the differentiated tokens represent different in-game collectables that come with different specifications proportional to their rarities (Figure~\ref{fig:F1}). 

\addtocounter{footnote}{-2}
\footnotetext{{\url{https://econ.st/31KV9w8}}}
\addtocounter{footnote}{1}
\footnotetext{{\url{https://www.f1deltatime.com/}}}
\addtocounter{footnote}{1}
\footnotetext{{\url{https://www.cryptokitties.co/kitty/896775}}}
Outside the conventional art industries, we also see participation in NFTs from the wider media community. As demonstrated in Figure~\ref{fig:alice}, The Economist, one of the most well-known magazines, minted its own NFT for its cover image of `Alice in Wonderland' featured with cryptocurrency \cite{the_economist_2021}. This is not an unprecedented move of the press industry, as other media outlets, including Times and The New York Times, have already jumped on the NFT bandwagon in the past~\footnote{\url{https://bit.ly/3kwanvA}}.
Examining the NFT market from users' angle, NFT trades appear to be clustered at the current phase: traders tend to focus on exchanging virtual objects of similar characteristics with others adopting similar specialisation \cite{nadini2021mapping}. The clustering phenomenon also takes place in the price determination of similar digital artworks \cite{fazli2021under}. Additionally, there appears to be some clustering for NFT artists too when it comes to their artworks' undertones: artists tend to exhibit clear emotional undertones with their digital artworks, where positive connotations are more common than negative ones \cite{franceschet2021sentiment}.


However, NFT is not a silver bullet to all the perplexities revolving around digital and media arts. First, it is worth questioning whether the explosion in the trade volume of NFTs truly remarks a reclaim of power by art creators through the decentralised technology, or it simply represents yet another asset bubble, which releases misleading signals that create a detrimental effect on the resale market due to oversupply of artworks \cite{franceschet2021crypto}. Apart from the debate of whether this financial inventiveness is de facto instrumental to artists \cite{zeilinger2018digital}, individuals should also stay cautious with the proclaimed technological robustness against fraudulence. It is worth noting that a wide-range implementation of NFT on digital arts is not equivalent to a termination of illegal appropriations: individuals can still take a screenshot of an artwork for personal or even illegal commercial use without being forced to pay the owner. Rather, as highlighted earlier in this section, NFTs should be thought of as licenses that outline the ownership. Although complete annihilation of illegal appropriations is unrealistic, we may find a silver lining in the composition of digital product valuation. In the context of customer value theory originated from marketing, how aesthetically appealing (emotional value) and the underlying social implication of owning the object (social value) may be more relevant to consumers of digital goods \cite{kim2011investigating}. Nonetheless, piracy and forgery can impede social value if digital arts are demanded to complement individuals' social status and the licensing itself is insufficient to secure the sense of exclusiveness. In addition, an NFT-backed virtual artwork is not tantamount to authenticity: there are several documented instances where malicious opportunists beguiled NFT collectors by selling digital artworks imposed as authentic pieces produced by famous artists \cite{Lee2021AllON}. Even worse, the possibility of scamming through `sleepminting'~\footnote{\url{https://bit.ly/3pPjF9j}} can bewilder inexperienced NFT collectors into considering a counterfeit as genuine. Therefore, one may come to the realisation that the non-fungibility feature of these digital tokens acts like a double-edged sword: we can only establish ownership by making the tokens differentiable, but it also implies they are not like cryptocurrencies that have going rates and are fungible. Consequently, authentication becomes more sophisticated for decision-makers.  Moreover, NFT does not grant users immunity to conventional scams such as impersonating as `customer service'~\footnote{\url{https://bit.ly/3pPy3P6}}. Thus, both artists and collectors should remain vigilant when conducting trades in digital artworks, albeit the remarkable progress the NFT stands for. Lastly, even when we temporarily drop the concern of NFT-related scams, the question of how precisely should an NFT-backed digital artwork be valued still lacks a commonly accepted conclusion. While some have reached out to the economics of supply and demand, which is ubiquitous in market analysis in countless industries, others have put forward more scenario-specific approaches. One instance is a rating system that values artworks based on key attributes, including past records of bids and sales \cite{franceschetenhancing}. Going beyond the discussion of the valuation of digital artworks, some have also inspected the accompanying trade mechanisms of NFTs through lenses such as game theory \cite{khezr2021property}. Apart from the NFTs' valuation, there are also rising concerns regarding the environmental implications associated with the large-scale adoption of NFTs. That is, the energy consumption that arises from minting and trading NFTs may potentially impede the process of achieving carbon neutrality. Nonetheless, businesses strive to ameliorate the situation by choosing production locations that use renewable energy and designing more efficient blockchains \cite{bruner_2021}.  In short, even if we omit the shortcomings of the NFTs in extinguishing ownership-related hitches, there still exists significant scope for further scrutiny in the practicality of NFTs.

In spite of the underlying challenges coupled with the utilisation of NFTs as a new means of accreditation for ownership, one should not easily dismiss its merits, especially when harnessed for benevolent uses with proper regulation. Specifically, a commonly accessible virtual trading system is restorative to self-employed content creators in virtual and digital art industries, both in the conventional and the emerging art industries that will be discussed in detail in this survey. Also, as aforementioned, the application of NFTs can extend beyond artworks in the absence of tangible physical forms. In theory, they can establish ownership for both the algorithm behinds AI-Dance/ Robotic Art or the entire artwork, both the physical and non-physical components. When it comes to the metaverse, the NFTs possess the ability to rise above mere ownership-verified collection in one's wallet by endowing digital artworks with additional interactive attributes. Thus, given the prominence of this burgeoning trading channel, the art and the wider community should seek to address its flaws and carefully scrutinise its possible prospect of delivering a socially desirable outcome, where the recipients of its benefits go beyond artists and curators.

\subsection{Contributions of the Survey}
As a concluding remark for the introductory part of this survey paper, we can obviously see the rising popularity of digital artworks, driven by Non-Fungible Tokens (NFTs). Nowadays, digital artworks have been widely traded via virtual platforms, i.e., the very early stage of the metaverse mainly motivated by economic efficacy. More importantly, the metaverse, as a socialised virtual world, can accommodate various artistic creation and subsequently facilitate virtual art trading. The work of art in virtual worlds can go beyond the fundamental functions of communication through images, sounds and stories. Also, the combination of the metaverse and computational arts could also serve as the key vehicle to encourage people from different cultures and across generations to express their opinions to each other in virtual worlds characterised by perpetuity and openness. 
We acknowledge that a broader definition of creators could include designers in other domains such as UI/UX for computer interfaces, interaction design, industrial products, and so on. Instead, we focus our review and discussion on the unique functions of computational arts in the metaverse, as defined in a prior survey~\cite{Lee2021AllON}.

We view the metaverse-driven worlds as an enormous and malleable canvas in the virtual-physical blended realities, while novel forms and representations of digital arts can appear in every aspect of our living environments once such realities are mature. 
With the background as aforementioned, our survey paper is a state-of-the-art review of computational arts, an interdisciplinary research area with arts and technologies. 
We did our utmost to provide comprehensive coverage in the metaverse era, such as virtual entities and scenes, calligraphy and poetry, robotic and immersive arts, and so on. 
More importantly, our survey serves as the first effort of integrated reviews for two complementary facets of computational arts and the metaverse.
To the best of our knowledge, the most relevant survey articles pinpoint either \textit{a focused area of computational arts}, e.g., art collection analysis and AI-generated images~\cite{james-she-survey} and media (video) intelligence~\cite{Guha2021ComputationalMI}, or \textit{fundamental developments of the metaverse}, e.g., technological enablers and ecosystems of the metaverse~\cite{Lee2021AllON}, the metaverse applications and industry sectors~\cite{industry-app-metaverse-sy}, the design of virtual worlds and metaverse user perception~\cite{metaverse-2013-csur}. In contrast, our survey paper aims to review the latest development of digital arts and investigate the potentials of computational arts in the metaverse era. 

\subsection{Structure of the Survey}
We review the latest development of various types of computational artworks, which could serve as the basic virtual building components in the metaverse, including computer-generated imagery like virtual photography and cinematic simulation (Section~\ref{sec:vpcs}), textual elements -- Calligraphy and Poetry (Section~\ref{sec:Calligraphy} and~\ref{sec:poetry}), and auditory and musical metacreation (Section~\ref{sec:auditory}). Next, the discussion focuses on human-engaged and user-centric artistic creation such as embodied collaboration (Section~\ref{sec:embodied}).
Then, virtual artworks represented by physical embodiments of robotics are discussed in Section~\ref{sec:robot}. 
Finally, we investigate the emerging multimedia of extended reality, known as Augmented Reality (AR) and Virtual Reality (VR), and the prominent features of virtual creativity and brainstorming that are highly relevant to the rise of the metaverse (Section~\ref{sec:ar-vr-art}).
After presenting the main body of article reviews, we discuss the research agenda of computational arts in the metaverse era (Section~\ref{sec:discussion}).

\section{Virtual Photography / Cinematic Simulation}\label{sec:vpcs}

In recent years, there have been a surge of artists working with digital virtual imagery, often inspired and influenced by the language of traditional photography and cinema, while exploring the new possibilities offered by computer rendering and simulation technologies \cite{paul-histories}. The broader context of this trend is that photography and cinema, the earliest forms of "new media" art, are becoming more computational and moving toward virtual worlds and networked spaces \cite{manovich2001language}. The boundaries between computer-rendered images and photography have been blurred \cite{koch20203d}. On the one hand, computer rendering is replacing traditional photography in many functional areas of application. When consumers today shop for a desk or visualise a new kitchen on 
Ikea's website, they may not realise that the images are likely to be computer renderings rather than traditional photography, yet this has long been the norm in the architectural and product design fields \cite{souppouris2012ikea} \cite{winchester2018putting}. These images replace the function of photographs for the lay viewer, at which point they become \textit{de facto} photographic images. As artist and professor Claudia Hart pointed out, computer generated imagery (CGI) is a form of post-photographic technology \cite{10.1215/17432197-1907190}. In this context, it also has become a new form of photography, namely virtual photography. One of the prominent examples of CGI as photography in popular culture is the virtual influencer \textit{@lilmiquela} on Instagram, a social media platform catered to image sharing. Taking on a persona of a virtual young female identity of Miquela, the account shares highlights in Miquela’s “life", embedding the realistically rendered virtual avatar into real environments, and often even "posing" with various real people and celebrities \cite{lilmiquela-about}. This is part of a recent trend, dubbed “celebrity 2.0”, of virtual stars created with CGI and simulation technology gaining popularity in the larger culture sphere \cite{10.1080/14680777.2020.1830927}.

\begin{figure}[!h]
     \centering
     \begin{subfigure}[b]{0.58\textwidth}
        	\centering
	\includegraphics[width=.49\linewidth]{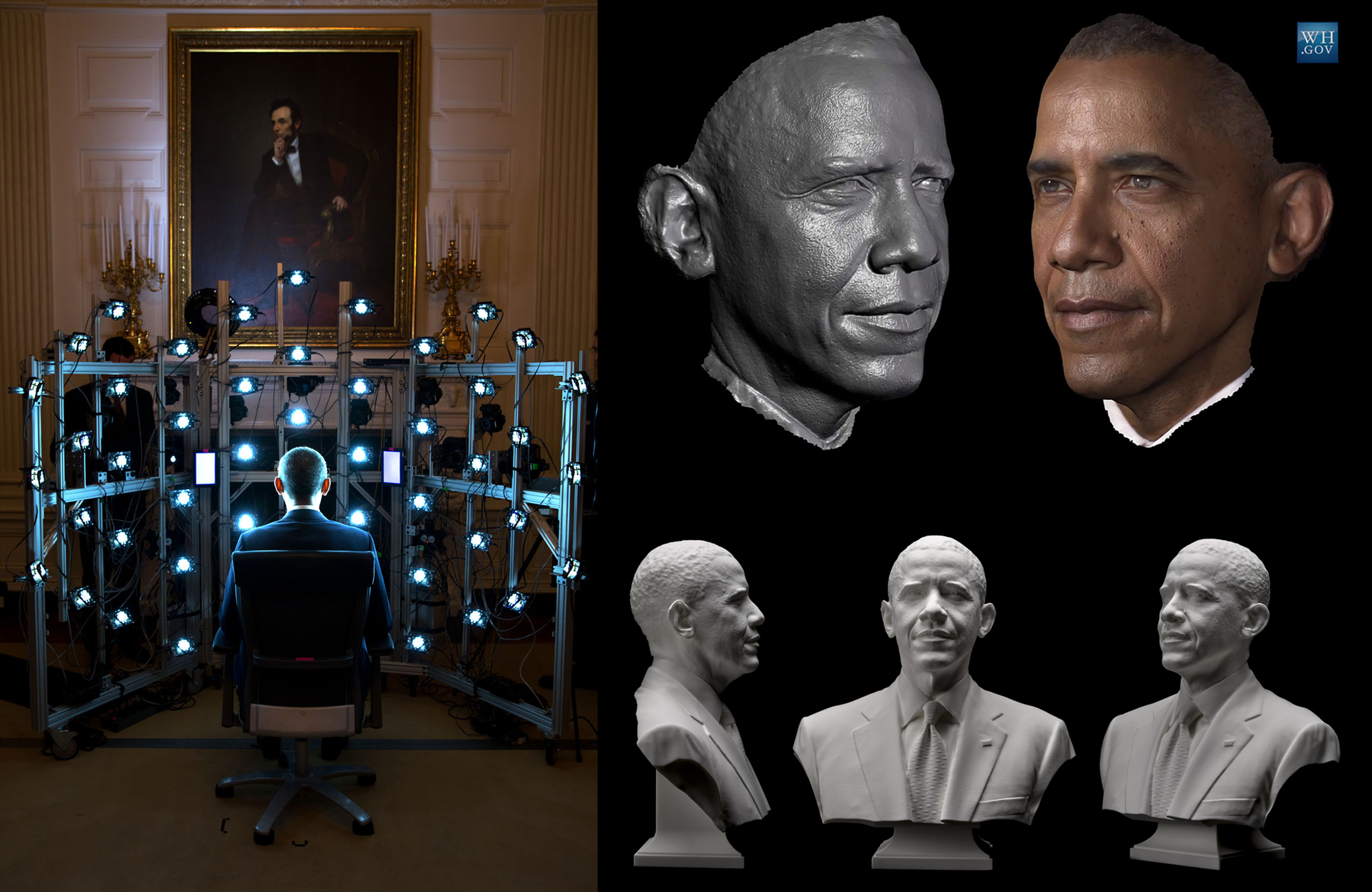}
	\caption{3D Portrait of former U.S. President Barack Obama, created with 3D scanning \& printing by 
	by the Smithsonian Institution.}
	\label{fig:obama-scan}
     \end{subfigure}
     \hfill
     \begin{subfigure}[b]{0.40\textwidth}
	\centering
	\includegraphics[width=.6\linewidth]{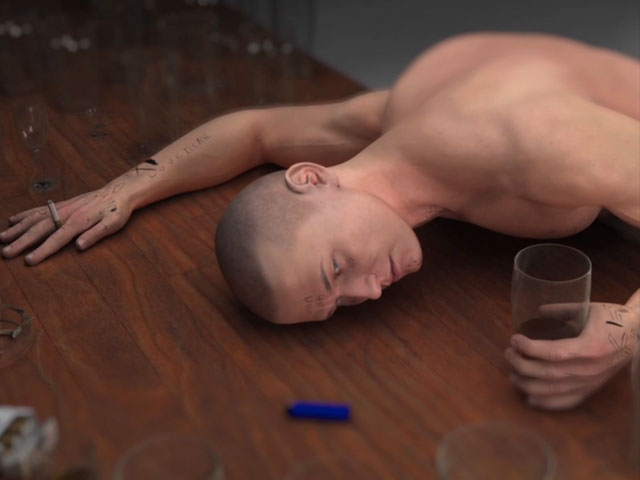}
	\caption{Ed Atkins. Still from \textit{Ribbons}, 2014, digital animation with sound, 13 mins, 18 secs.}
	\label{fig:atkins-ribbons}
     \end{subfigure}
        \caption{Virtual Characters}
        \label{fig:v-character}
\end{figure}

On the other hand, a variety of new visual sensing technologies, homologous to traditional lens-based photography, are expanding new realms of application. Perhaps one of the best known of these applications in popular culture includes the fact that in 2016, former U.S. President Barack Obama became the first president to have an official 3D portrait created by a team led by the Smithsonian Institution, through a process that involved capturing facial form, colour and texture with a mobile Light Stage equipped with fourteen cameras and fifty lights. Combined with data of the rest of the head and shoulders captured by a structured light 3D scanner, the Smithsonian team was able to recreate a complete and accurate 3D representation with which to produce the physical bust using 3D printing \cite{kristy2014president} (Figure~\ref{fig:obama-scan}). In many ways, 3D scanning shares or utilises some of the same techniques as photography, capturing traditional visual information with a camera while using enhanced assistive means, such as structured light pattern or photogrametry, to capture additional spatial distance information to not only record colour and light, but also the three-dimensional forms \cite{10.1145/1667239.1667247} \cite{830281592}. In this post-photographic era, 3D scanning can be seen as a form of photography in the broadest sense, or Augmented Photography. This technique has in recent years been extensively studied in the museum field and is being applied to cultural heritage preservation, with the hope that this new augmented photography will not only provide a 2D record as traditional photography does, but also a complete record of the three-dimensional morphological details of the artefacts with surface material and colour properties, thus enabling reconstruction in the three-dimensional space \cite{10.1016/j.culher.2006.10.007} \cite{10.1179/019713609804516992}.


In addition to still images, CGI can naturally be used to create moving images. In the early days of CGI, this field was only accessible to animation studios or teams because of the time and labor involved, and in the personal creation and research domain, artists often need to collaborate with technology research teams or gain access to sophisticated computing facilities \cite{9780500203989}, such as the case of Rebecca Allen using early 3D digitization and animation technology to create a music video for the pioneering German electronic music group Kraftwerk in 1986 \cite{watson-rebecca}. Or in another case, the entire creation is driven by people who are engineers by profession, such as the simulated creatures produced by engineer and visual effects developer Karl Sims, which are exhibited as works of art while winning the Prix Ars Electronica media art award in both 1991 and 1992 \cite{sims-website}. With the increased usability of software, many artists have begun to adopt CGI for personal artistic expression. One notable recent example is British artist Ed Atkins, who uses CGI to create highly personal video artworks, which is noted for its “syncopated montages of sounds and filmic images. \cite{26355345}” His works, such as  \textit{Ribbons} (2014) (Figure~\ref{fig:atkins-ribbons}), often show virtual male characters speaking poetic or mundane phrases with ambiguous meanings, accompanied by slightly exaggerated camera movements, bokeh, lighting and dust effects, creating a rich and layered sense of humour and melancholy at the same time. When it comes to the techniques used, Atkins believes that animating with virtual avatars allows him to work as an independent artist and to blend visual, sound and narrative elements. The avatars allow the artist to take on different persona and express his thoughts and emotions without constraint~\cite{atkins-interview}. With emerging tools targeting virtual human creation such as Metahuman and detailed 3D-scan libraries such as Quixel Megascans, high-quality CGI creation will become even more accessible to creators.

\begin{figure}[!h]
     \centering
     \begin{subfigure}[b]{0.3\textwidth}
     	\centering
	\includegraphics[width=\linewidth]{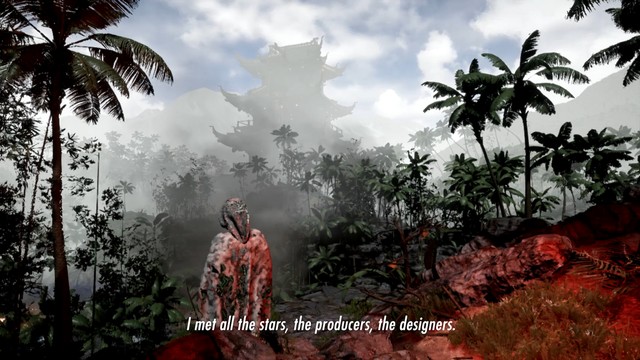}
	\caption{Lawrence Lek. Still from \textit{AIDOL}, 2019, digital animation with sound, 83 minutes.}
		\label{fig:lek-aidol}
     \end{subfigure}
     \hfill
     \begin{subfigure}[b]{0.3\textwidth}
		\centering
	\includegraphics[width=.788\linewidth]{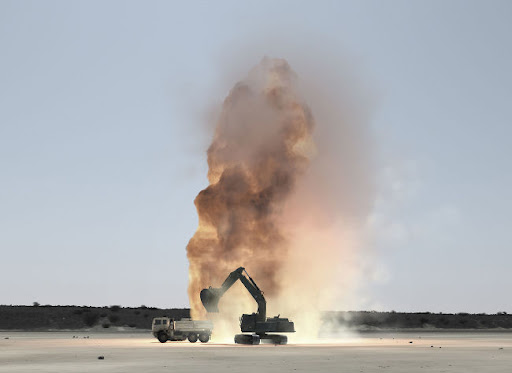}
	\caption{John Gerrard. Still from \textit{Live Fire Exercise (Djibouti)}, 2011, simulation.}
	\label{fig:gerrard-live}
     \end{subfigure}
     \hfill
     \begin{subfigure}[b]{0.3\textwidth}
	\centering
	\includegraphics[width=\linewidth]{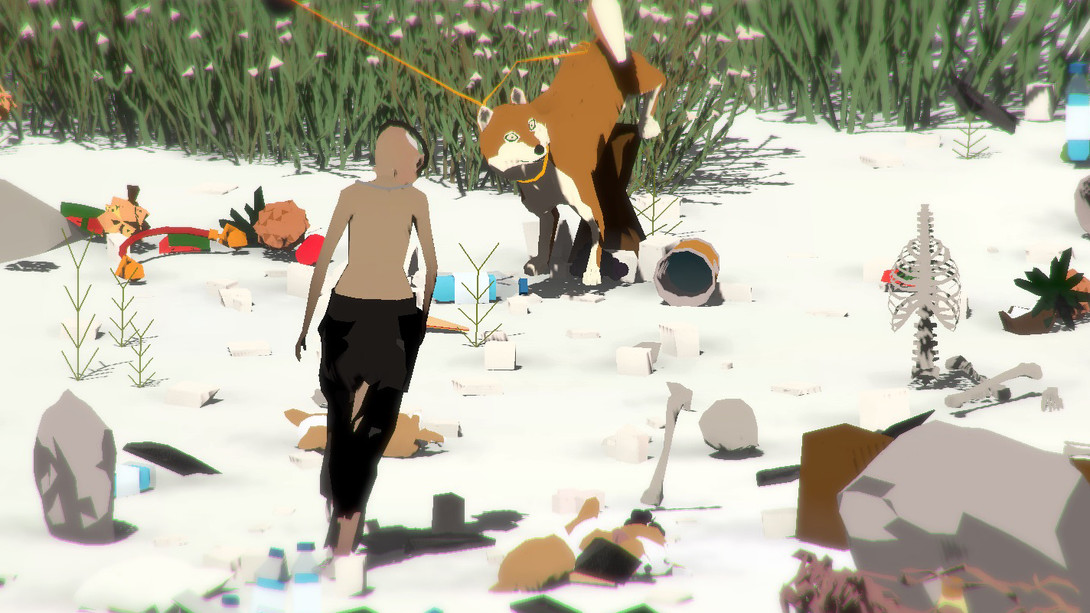}
	\caption{Ian Cheng. Still from \textit{Emissary Forks At Perfection}, 2015-2016, live simulation.}
	\label{fig:cheng-emissary}
     \end{subfigure}
        \caption{Virtual Scenes and Environments}
        \label{fig:v-environment}
\end{figure}

The development of 3D graphics in the field of games is also encouraging. Improvements in usability and graphics quality in game engines have made it possible for cultural and artistic creators to use them as a regular tool \cite{10.1007/978-3-030-14132-5_16}. Game engines have the benefit of real-time rendering compared to traditional 3D content creation tools, allowing artists to get visual feedback immediately without having to spend time waiting for rendering results, making the workflow more intuitive and smooth. At the same time, the "virtual production" pipeline based on game engines also enables film production crew to quickly obtain the virtual backgrounds needed for visual effects, which can be shown on large LED screens on set in the studio and captured directly by the camera at the same time as the performances, without the need for complex and troublesome green screen shooting and compositing in post-production \cite{martinez2021virtual}. This has become one of the new trends worth exploring in the field of video production. In the realm of using game engines to create moving-image artworks, Lawrence Lek, an artist with a transnational background, uses Unreal Engine to produce art films with science fiction narratives \cite{lek-website}, which has been noted for his exploration of the concept “Sinofuturism” \cite{10.5749/vergstudglobasia.7.2.0086}. Lek’s narrative themes span artificial intelligence, financial market, simulation, and the identities of Singapore and Asia. In several of Lek’s works such as \textit{Geomancer} (2017) and  \textit{AIDOL} (2019) (Figure~\ref{fig:lek-aidol}), the protagonists are either an artificial intelligence who want to break out of their constraints and to create art, or a human creator whose relevance has faded due to the popularity of “creative AI,” thus highlighting the tension between AI and creativity. The films present large-scale futuristic sci-fi cityscapes, and in about a decade ago, such works would have been almost impossible to be done by a single artist due to software usability and  the limitations of computing power. Today, thanks to the relative ease of use of game engines and the continuous improvement of real-time rendering quality, artists can avoid the time and expense associated with CGI and complete long-form video works that emphasise conceptual thinking and narrative expression in a relatively short period of time.


In addition to its real-time rendering capability, the game engine also allows for programmed mechanism and interactivity due to its computational nature, which means that the generated image no longer has to be on a fixed timeline, but instead can be full of unexpected dynamic events. Artworks created by taking advantage of this feature can be considered as "artistic simulation," when the image progresses according computation and is visually represented through time. When such artistic expression includes influences from the language of traditional cinema, such as camera movements and editing techniques, it can be also considered as "Cinematic Simulation." The ideas behind these terms have been practised by artists actively in recent years and discussed in art reviews when referring specifically to the works of these artists, but in-depth theoretical treatments on simulation as art remain sparse \cite{27992174}. One prominent example of practice is the British artist John Gerrard, who uses the game engine Unigine to create surreal simulated landscapes (Figure~\ref{fig:gerrard-live}). Gerrard states that while there is always the concept of duration in traditional cinema, his work, though still visually presented as video, is theoretically endless because of the use of real-time simulation techniques \cite{gerrard2016interview}. This idea is best reflected in one of his early works, \textit{One Thousand Year Dawn} (2005), in which the simplistic image shows a character standing on a beach facing a sunrise that is set to take a thousand years to complete. Many of Gerrard's recent works feature meticulously reconstructed virtual representation of industrial and agricultural facilities, such as data centers, solar power plants, and pig farms, often in remote and uninhabited locations. Gerrard "simulates" the form and dynamics of these facilities, while the time-of-day and weather in the images are synchronised with the location of the facility. He conceptualises these works as digital portraits of such facilities. During the production process, the artist's team took thousands of photographs in the fields around the facilities, which were then used to reconstruct the sites as detailed virtual environments. In this sense, Gerrard's work can be considered an example of the exploration of the concept of the digital twin within the field of art.


Another notable artist using simulation technology is the New-York-based Chinese-American artist Ian Cheng. Inspired by his educational background in cognitive science, Cheng's work explores the agency and interactivity of virtual intelligence using “live simulation,” or “video games that play themselves” \cite{scott2017watch}. The work \textit{BOB (Bag of Beliefs)} (2018-2019) is a snake-like monster lifeform that is alternately driven by multiple sub-agent systems with different personalities and desires, generating corresponding judgements, speculations, reactions and memories to various events and elements that appear in the environment in real time. Meanwhile, viewers can interact with the snake monster through their cell phones and send virtual "offerings" and text messages to the snake monster, thus influencing its response to the environment and objects. In this process, new dynamics are generated between different viewers' behaviours, so that new relationships are formed between viewers through the connection of the snake monster. Cheng considers his work as a construction of a kind of “Minimum Viable Sentience” \cite{cheng2020minimum}, and “BOB” is indeed an experiment on the interaction between intelligent systems and people in the context of art. Ian Cheng's other series, \textit{Emissary Trilogy} (Figure~\ref{fig:cheng-emissary}), is also an exploration of artificial intelligence systems in the form of simulation presented as moving images. In each of the trilogy's environments, multiple AI characters with different goals interact with each other, driving new events in the virtual world and propelling the narrative forward. This is arguably a new way of making narrative moving-images. There is no need to write a script in advance, whether by a human being or AI. Everything is set up so that the AI agents will act according to their respective agenda, and then the story is formed in a generative way through events triggered by the evolution and interaction of the AI agents. This form of work can be considered as "event-driven cinema".

\begin{figure}[!h]
     \centering
     \begin{subfigure}[b]{0.5\textwidth}
        	\centering
	\includegraphics[width=\linewidth]{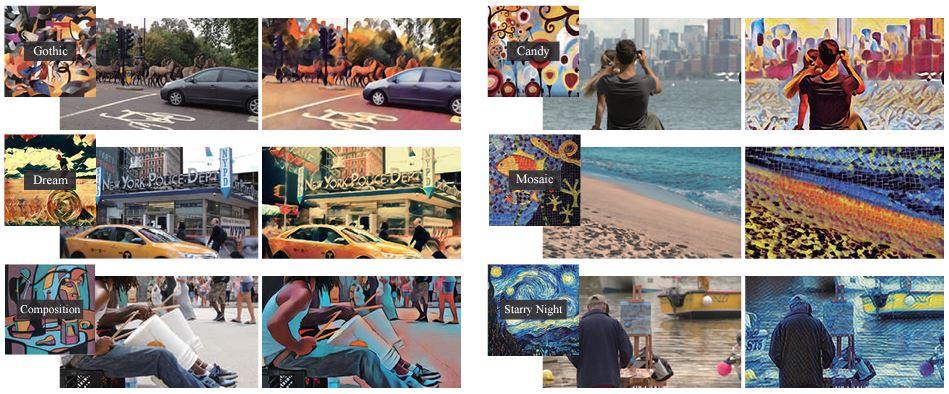}
	\caption{Style transition with multiple styles.}
	\label{fig:cccss-vs}
     \end{subfigure}
     \hfill
     \begin{subfigure}[b]{0.48\textwidth}
	\centering
	\includegraphics[width=\linewidth]{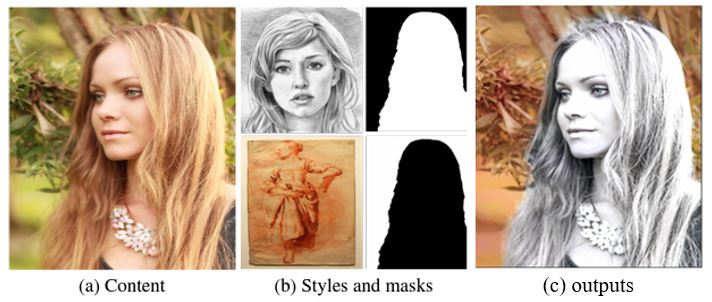}
	\caption{Segmenting the background and portrait.}
	\label{fig:v-ss}
     \end{subfigure}
        \caption{Examples of Style Transfers, applicable to both virtual environments and physical worlds.}
        \label{fig:transfer-tras2}
\end{figure}
Before we discuss other building blocks for metaverse composition, we highlight the potentials of artistic augmentation of computer-generated virtual scenes and characters. The latest techniques of neural style transfers can achieve real-time processing of videos. 
The videos of virtual or physical worlds can be alternated by multitudinous user-defined styles~\cite{realtime-style-transfer} (Figure~\ref{fig:cccss-vs}), which achieve high reasonable granularity of processing the video scenes by providing multiple styles for different object classes~\cite{Xia2021RealtimeLP-segment} (Figure~\ref{fig:v-ss}). 
This implies that metaverse creators can choose their preferred styles, and offer highly personalised yet artistic experiences to other metaverse individuals, due to the nature of cinema-alike metaverse. 

\section{Calligraphy} \label{sec:Calligraphy} 


As discussed in the previous section, images and videos (e.g., painting) are excellent candidates to describe and illustrate abstract concepts. In contrast, textual elements are indispensable in the cyberspace of virtual worlds, which serve as crucial containers to express ideas and support user communications in accurate and efficient manners. 
The existing metaverse virtual worlds, such as \textit{Fortnite}\footnote{\url{https://www.epicgames.com/fortnite/en-US/home}}, allow users to select various skins and items to customise their avatars. Such avatar makers reflect the users' desire to personalise their virtual characters and create unique representations. 
In a similar way, textual contents in the metaverse can be revealed as dialogues between virtual characters and interaction footprints on virtual objects (e.g., writings on a virtual wall) in the metaverse~\cite{DBLP:conf/mm/KumarBL021}. However, the current capabilities of mobile interfaces for user interaction with virtual environments suffer from throughput rates and levels of input details~\cite{csur-lee2022}. 
For example, the current input devices may not be able to support virtual pens or brushes for artistic and personalised handwriting in virtual 3D worlds, with full granularity, such as stroke width and legibility. 
Also, writing calligraphy would be more complicated than the counterpart of building avatars. Instead of requiring intensive training in calligraphy, every participant in the metaverse should enjoy the democratisation of artistic creation~\cite{Lee2021AllON}.
Thus, metaverse users would need alternative platforms and tools for customising their texts with artistic effects. 

\begin{figure}[!h]
     \centering
     \begin{subfigure}[b]{0.54\textwidth}
	\centering
	\includegraphics[width=\linewidth]{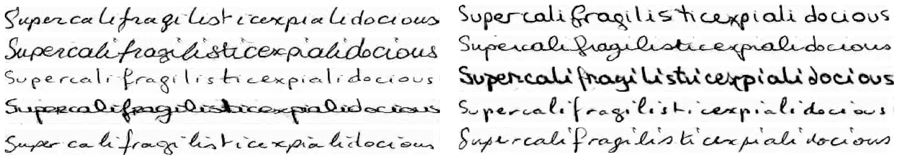}
	\caption{Ten different styles of English calligraphy from ScrabbleGAN~\cite{Fogel2020ScrabbleGANSV}.}
	\label{fig:34char-word}
     \end{subfigure}
     \hfill
     \begin{subfigure}[b]{0.44\textwidth}
	\centering
	\includegraphics[width=.68\linewidth]{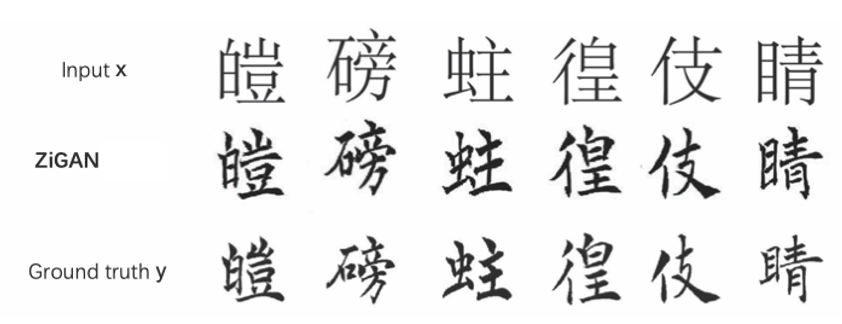}
	\caption{Style transfer for Chinese calligraphy from Zigan~\cite{10.1145/3474085.3475225}.}
	\label{fig:zigan}
     \end{subfigure}
        \caption{Calligraphy in English (Left) and Chinese (Right).}
        \label{fig:calligraphy}
\end{figure}

Nowadays, artificial intelligence (AI) can recognise the features of calligraphy and result in promising effects of mimicking experts' calligraphy~\cite{Zhao2020DeepIH}. 
Considering that the metaverse has to serve human users from highly diversified backgrounds, the existing AI techniques can quickly adopt multiple linguistic contexts.
As demonstrated by GANWriting~\cite{Kang2020GANwritingCG}, AI-generated handwriting can mimic various features of handwriting style, including roundness, stroke width, skew, ligatures, and slant. Such generated handwriting texts are highly realistic, where human evaluators cannot judge whether machines or human beings write it. 
In addition, ScrabbleGAN~\cite{Fogel2020ScrabbleGANSV} can synthesise handwritten English text of varying lengths, which concatenates character-token of unlabelled data into word-level calligraphy of various styles, for instance, ``\textit{Supercalifragilisticexpialidocious}'' from Mary Poppins (Figure~\ref{fig:34char-word}). Accordingly, HiGAN~\cite{Gan2021HiGANHI} significantly improves the similarity with the specific reference of handwriting style, 
while JokerGAN~\cite{Zdenek2021JokerGANMM} enhances the text line conditioning for more natural and intuitive handwriting.

In the context of Asian calligraphy, a most recent model of few-shot style transfer, 
namely ZiGAN~\cite{10.1145/3474085.3475225}, demonstrates fast learning of Chinese calligraphy without significant sample numbers of the handwriting of calligraphy masters that are hard to acquire in practical scenarios (Figure~\ref{fig:zigan}). In addition, the calligraphy artwork can further consider the emotions of metaverse users, by detecting users' emotions. 
For instance, facial expression and heartbeat rates can be supported by computer vision and smart wristbands, respectively. In ECA-GAN~\cite{calli-emotions}, the calligraphy style can employ various emotions, including peaceful, happy, and sad, to adjust the structures of Chinese characters in both character and discourses levels.
Furthermore, the technique of photometric stereo can further convert Chinese calligraphy into 3D virtual objects. Remarkably, the heightmap of the reconstructed 3D surface can present an enriched digital copy pinpointing the beauty of Chinese calligraphy, in terms of strength and personality~\cite{Jian2020LearningTT}. This opens the opportunities of circulating artworks as 3D virtual objects made by calligraphists in the metaverse.





\section{Poetry}\label{sec:poetry}
The above paragraphs describe the properties of AI-driven artistic representation of characters, and the potential benefits of enriching the metaverse. The latest technology of automatically generated poems, such as \textit{Deep-speare}, can create Shakespeare's poems that are hardly distinguished by human beings~\cite{9078455}.
Furthermore, the existing computational approaches can generate impressive linguistic contents for such artistic texts (e.g., poems\cite{Ghazvininejad2016GeneratingTP, 2018A, DBLP:conf/aaai/DengWLCXZWX20}). 
Accordingly, the union of AI-generated calligraphy and poetry genres can facilitate the linguistic creativity of creators and artists in various scenes inside the virtual-physical blended metaverse. 
More specifically, the prior studies imply that activities between metaverse users, i.e., user-generated content including discussion topics and scenes, can achieve augmentation linguistically and artistically. 
Recurrent neural networks (RNN) can process discussion topics and related words, resulting in automatic poetry generation~\cite{Ghazvininejad2016GeneratingTP, 2018Stylistic}. 


\begin{figure}[!h]
     \centering
     \begin{subfigure}[b]{0.51\textwidth}
	\centering
	\includegraphics[width=\linewidth]{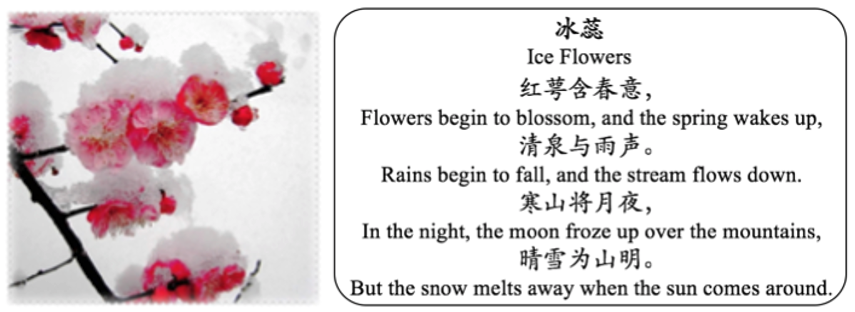}
	\caption{An example of five-character quatrains, a type of Chinese poems~\cite{2018A}.}
	\label{fig:hieas2s}
     \end{subfigure}
     \hfill
     \begin{subfigure}[b]{0.48\textwidth}
	\centering
	\includegraphics[width=.79\linewidth]{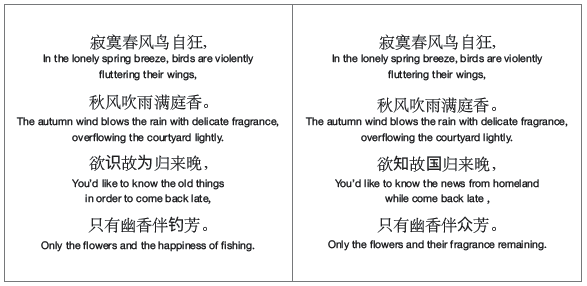}
	\caption{Iterative polishing a seven-character quatrain: Original (Left) and Right (Polished)~\cite{DBLP:conf/aaai/DengWLCXZWX20}.}
	\label{fig:polish-poem}
     \end{subfigure}
        \caption{Machine-generated poems that leverage pictures and topics to create quatrains.}
        \label{fig:peom}
\end{figure}

Moreover, metaverse users can leverage poetry as a channel of two-way communication with gamified elements in virtual worlds, as reflected by earlier examples of human-to-human (Super Atari Poetry~\cite{10.1145/1459359.1459602}) and human-to-AI (ReadingRites~\cite{10.1145/3325480.3329182}) games through the existing cyberspace. 
It is also important to note that AI-driven linguistic creation demonstrates the potentials to match virtual scenes in VR and physical environments in AR. This connects to an analogy to the fusion among calligraphy, poetry and painting, highly influenced by the philosophical and spiritual tradition in ancient China, namely, Taoism and Confucianism.
AI can recognise scenes (input images), and accordingly generate various poetry genres, with the following examples.

A hierarchy-attention seq2seq model, namely HieAS2S~\cite{2018A}, can generate two types of Chinese poetry genres: five-character quatrains (\textit{WuJue}) and seven-character quatrains (\textit{QiJue}), based on the image recognition module supported by `\textit{GoogLeNet}'\footnote{\url{https://pytorch.org/hub/pytorch_vision_googlenet/}} to understand a scene describing `red plum' and `snow' (Figure~\ref{fig:hieas2s}). 
In the context of AI-generated calligraphy, we judge the effectiveness by the level of intimacy to the human handwriting. Similarly, we should also evaluate AI-generated poetry through its fluency, meaningfulness, phonological compliance, and coherence~\cite{8949208,8482485}. 
As such, Deng et al.~\cite{DBLP:conf/aaai/DengWLCXZWX20} attempt to narrow down the performance between poetry generated by human beings and machines. In their work, Quality-Aware Masked Language Model drives the iterative polish of quatrain by considering the tokens, segmentation, positions, tones, and rhymes in AI-generated poems (Figure~\ref{fig:polish-poem}). Other works show the feasibility of automatically generated poetry across age and culture, e.g., Tang-dynasty poetry~\cite{8482485}, Modern Chinese poetry~\cite{9206888}, Spanish poetry~\cite{10.1145/3078081.3078085}, as well as Homeric poetry~\cite{8940426}. 

Although the existing studies show promising performance in the scene-poetry automatic generations, the grand challenges are as follows. 
First, poetry creations have to maintain consistency between the semantics of poems and images. We have to understand the reasons for topic drift throughout the automatic generation process, and to design preventive mechanisms, perhaps more transparent and explainable tools are required. Also, some words appear more frequently than the rest, but they could deteriorate the model effectiveness~\cite{10.1145/3381858}. 
On the other hand, we would like to emphasise that most automatically generated poetry are considered as rigid outputs. Recalling an earlier example (Figure~\ref{fig:hieas2s}), the augmentation of five-character quatrains on top of a physical scene, a type of Chinese poem, is a potential scenario of art applications~\cite{2018A}. 
We expect that the metaverse would serve as a huge creation space that can accommodate flexible and free artwork representations, 
known as `\textit{visual poems}'. For instance, some visual poems\footnote{\url{https://www.michaelandsarachaney.com/blog/2018/11/21/ten-great-examples-of-visual-poetry}} explore the highly blended representation between visuals and textual contents. 
Significant research efforts can focus on the spatial factors in virtual and physical worlds and even virtual-physical reality scenarios.

\section{Auditory and Musical Metacreation}\label{sec:auditory}

\begin{figure}[!h]
     \centering
 \begin{subfigure}[b]{0.3\textwidth}
     \centering
\includegraphics[width=\linewidth]{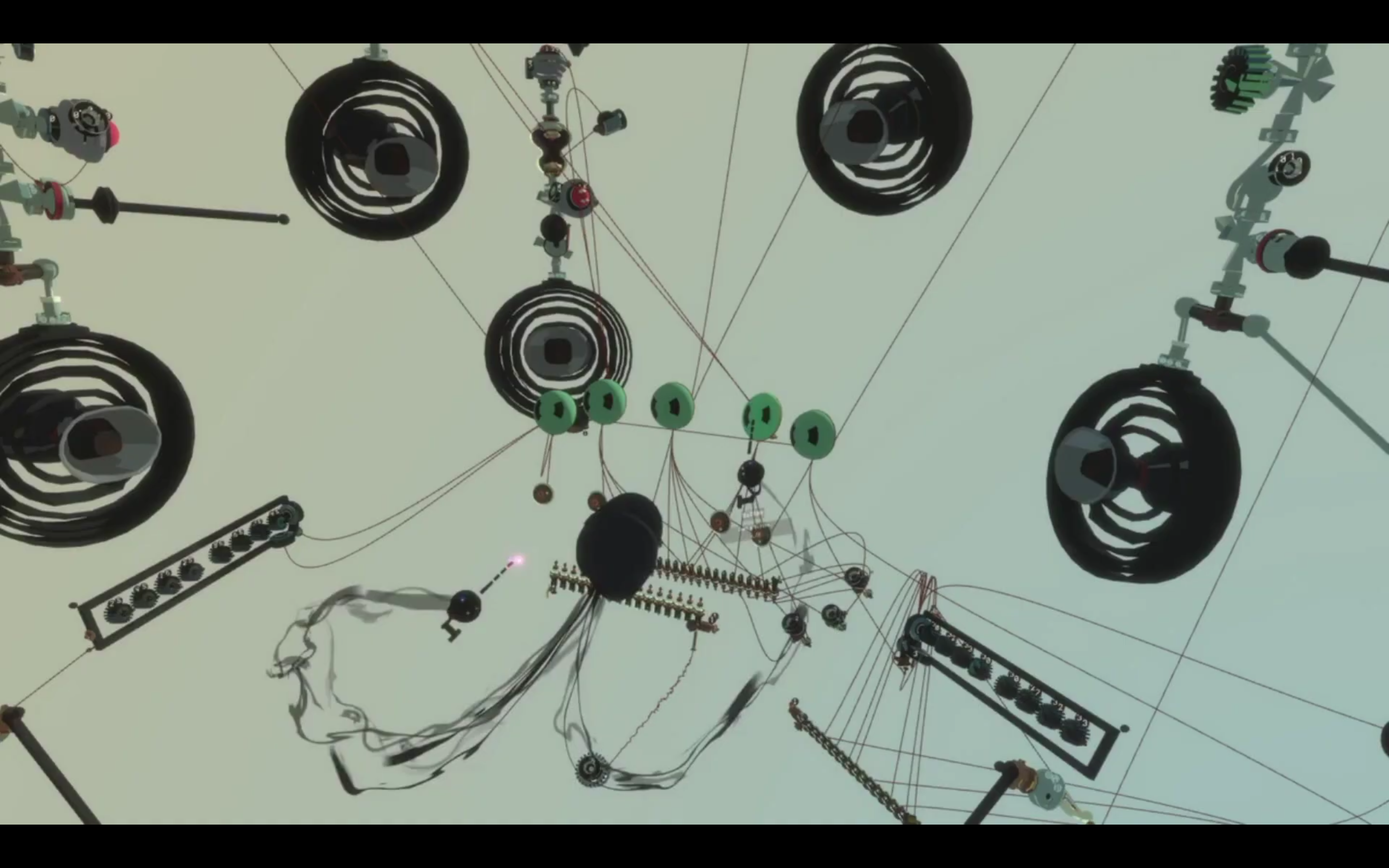}
	\caption{PatchXR, a VR musical instrument, turning a room into music studio on the fly \cite{patxr2020}.}
	\label{fig:PatchXR}
     \end{subfigure}
     \hfill
     \begin{subfigure}[b]{0.3\textwidth}
   \centering
	\includegraphics[width=.9\linewidth]{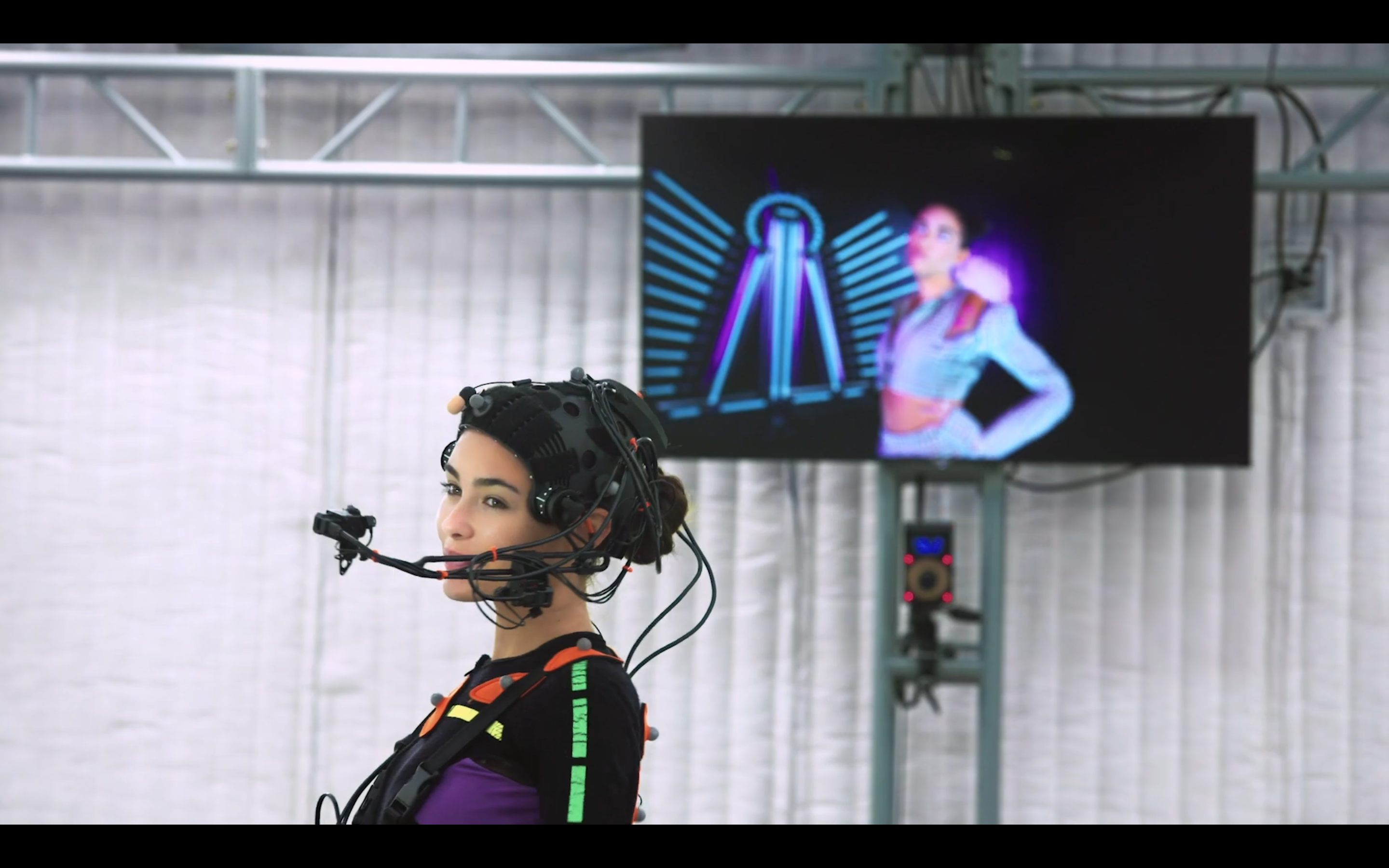}
	\caption{Singer Madison Beer performing in a studio, while audiences watching performance via her avatar in a virtual world \cite{madison2020}.}
	\label{fig:madison}
     \end{subfigure}
     \hfill
     \begin{subfigure}[b]{0.3\textwidth}
         \centering
	\includegraphics[width=\linewidth]{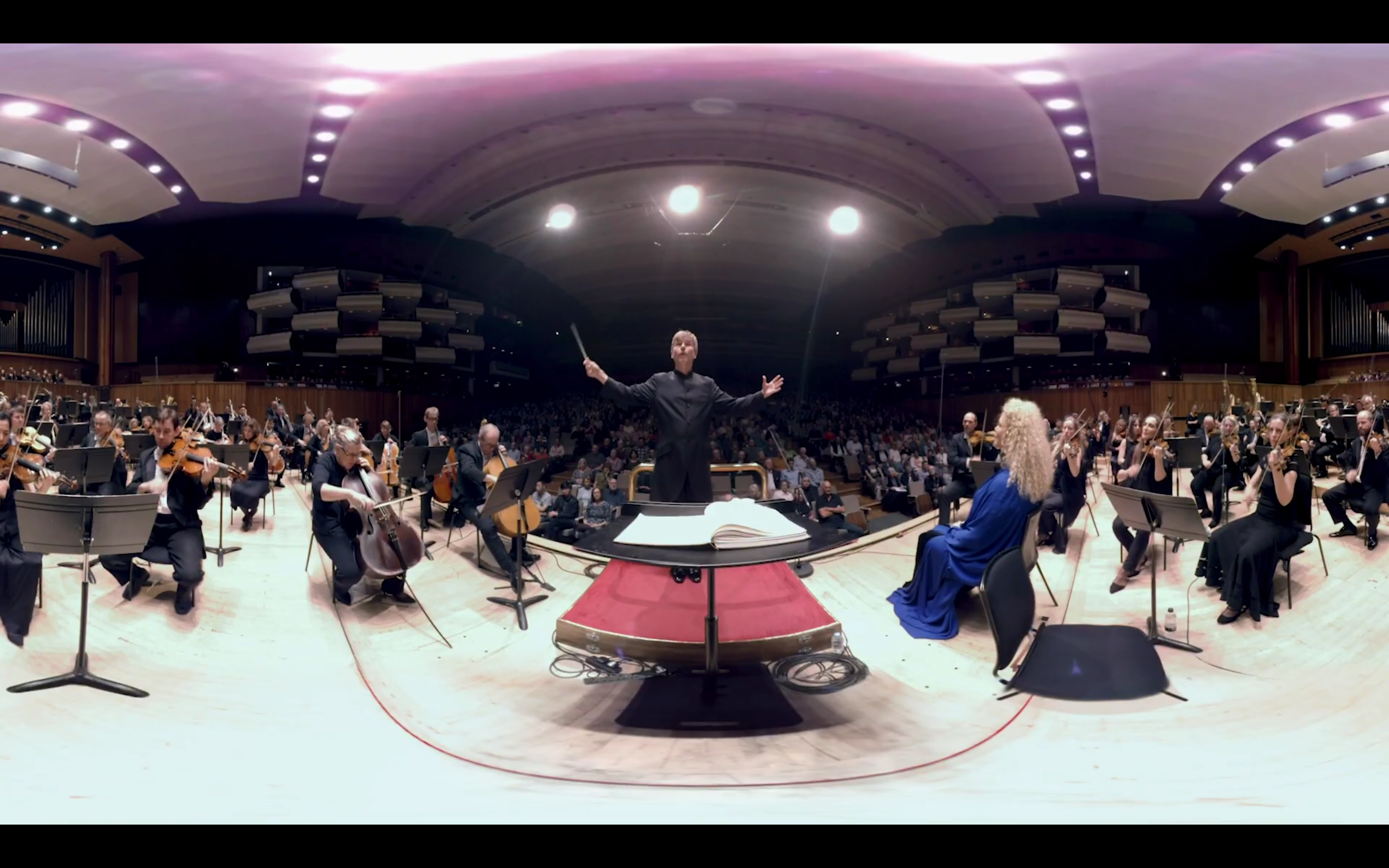}
	\caption{360° Egocentric view of the audience watching orchestra in the virtual space \cite{orchestra2020}.}
	\label{fig:orchestra}
     \end{subfigure}
        \caption{Musical Arts in the metaverse.}
        \label{fig:data-art}
\end{figure}

\paragraph{Musical Instruments for the Metaverse}
In 1932, Antonin Artaud published \emph{Theatre of cruelty} in which he foresaw the emergence of the metaverse and in there, recognized the unique role of musical instruments in this space \cite{artaud2018theatre}. Fast forward 60 years, in 1992, Jaron Lanier, one of the founding fathers of Silicon Valley, gave a presentation titled \emph{Sound of One Hand} at the SIGGRAPH Conference, where he demonstrated how the sound was being synthesized and improvised in virtual reality. This demonstration marked the beginning of the musical instruments being created entirely in virtual space. After almost 20 years of stagnancy in the field, an audiovisual environment for audio composition and performance, named \emph{Versum} was presented in 2009 \cite{lTarikBarriVersumDemo}. It features audiovisual objects arranged in virtual space in which the composer, performer, and spectator can float freely. It uses virtual space as a canvas for composition, which is explored by different paths so each run creates a meta-composition. A virtual reality musical instrument (VRMI), \emph{Carillon} was presented in 2015 by Robert Hamilton and Chris Platz which allows users immersed in VR to control using gesture and motion\cite{Carillon2015}. In 2018, he presented \emph{Cortet}, another VRMI which can mimic well instruments like piano, or percussion very well \cite{Coretet2018}. It allows multiple performers to collaborate in a shared virtual space. In 2019, \emph{Connexion} was presented by Danowski et al., which embedded the idea of magical and mimetic, i.e. an instrument in a quasi-organic form that is visually responsive to the performer's actions \cite{Connexion2022, paul-panopticon}. It surrounds the audience with an eight-channel sound system, making the audience immersed from every direction. Most recently, \emph{PatchXR} has taken a major leap by providing a spatial equivalent of visual programming engines, which allows artists to build their audiovisual instruments using patches, i.e. visual programs which represent the flow of events in the music \cite{patxr2020}. It allows artists to turn a place into a music studio on the fly, to compose and play music in a modular and sculptural workflow with the building blocks of sound at the spatial level (Figure \ref{fig:PatchXR}).

\paragraph{Music Concerts and Festivals in the Metaverse}\label{concert}
Nowadays many VR platforms like VRChat, AltspaceVR allow artists to create their world and perform in the virtual environment together, and some of these platforms have millions of users around the world as a potential audience. Given the severe impact on the live-music performance industry by the coronavirus pandemic, the virtual world has emerged as an ideal stage for music concerts and festivals. VRChat platform has been used for the live concert with support from engineers from VRrOOm since 2020 \cite{jarre2020}. In 2020, Sony organised a live concert of musician Madison Beer. As shown in Figure \ref{fig:madison}, while she stood in a studio in her virtual reality suit, the audience received the appearance of a live concert \cite{madison2020}. Some VR platforms like the Fortnite game platform have also been used to the concert by playing back animation inside the gaming engine, instead of the musician performing live \cite{scott2020}. Minecraft, another VR gaming platform, has been used to organise music and film festivals \cite{vrevents2022}. One of the major advantages of organising concerts and orchestras in the virtual space is that it allows the audience an opportunity to know their inner workings. For example, during Philharmonia Orchestra conducted by Esa-Pekka Salonen (Figure \ref{fig:orchestra}), the audience was able to move between each instrumental group, go backstage, stand next to the conductor, or watch from the theatre seats, which is nearly impossible during such events in the real world due to maintenance and security costs \cite{orchestra2020}. Moreover, conducting concerts in the metaverse offers performers to bring a new perspective of decorating the stage environment as per the theme of their performance, as in Alice in Wonderland. 2020 Burning Man event was organised in the multiverse using the Altspace VR platform \cite{burningman2022}. Not to mention, concerts in the metaverse will also provide easy accessibility to physically disabled audiences who often find it challenging to attend real-life concerts.

Though it has generated interest among the audience who are VR users, the technology is still in the nascent stage for organising a concert for a large audience. For example, VRChat allows only a maximum of 40 users in one room. So, even though multiple instances of concert rooms could be created to accommodate a large audience, the avatar of the performer could be present in only one room. Other challenges include audience jumping to restricted areas which can block the projection of the artist's avatar. The major challenge lies in how to simulate the realistic concert environment since the real concert tends to have lots of chaos in terms of the auditory environment, i.e. main music coupled with sounds from the live fans, and other background noises. One possible approach is to sample sound from different areas of a concert area which allows the sound to be activated in an even or uneven manner in six degrees of freedom within the sound field.

\paragraph{Musical Creativity for the Metaverse}
While replacing the physical auditorium with a single website is one major key focus for the metaverse, another focus also lies in generating musical arts itself with help of machines that can assist music concerts in the metaverse. The field of Musical Metacreation (MuMe) especially focuses on this goal. Metacreation is the idea of endowing machines with creating behaviours that are deemed to be novel and creative by unbiased human judges \cite{whitelaw2004metacreation}. According to Pasquier et al. \cite{pasquier2017introduction}, the MuMe field is about developing software and systems that can autonomously (or interactively) recognise, learn, represent, complete, accompany, compose, or interpret music. As mentioned in the previous paragraphs, 
though concerts in the metaverse give the appearance of live performance, it is still far from simulating a real concert-like experience. Music metacreation can fill this missing gap, such as machines generating the background sound depending on the concert theme, concert hall settings, and audience crowd in order to stimulate the chaos/adrenaline rush feelings of a real concert. 

Recently deep learning-based approaches have taken the central stage in human-machine collaboration for music generation. These approaches can be divided into two categories based on their input method, whether they use either note sequences or raw audio as their input. Magenta is an open-source music project from Google that uses a regular recurrent neural network (RNN) and two Long Short-Term Memory (LSTM) \cite{roberts2019magenta}. It can handle any Musical Instrument Digital Interface (MIDI) files. BachBot\footnote{\url{https://github.com/feynmanliang/bachbot/}} is LSTM  based approach that aims to generate and harmonize chorales indistinguishable from human-made chorales. It is considered one of the best efforts in handling polyphonic music as the algorithm can handle up to four voices. FlowMachines is a commercial project from Sony which uses Markov constraints as neural network techniques. It has generated the first AI pop-songs \cite{flowmachines2016}. Wavenet\footnote{\url{https://magenta.tensorflow.org/2016/09/23/learning-music-from-learned-music/}} is another project from Google based on Convolutional Neural Networks (CNN) which aims to enhance text-to-speech applications by generating a more natural flow in vocal sound \cite{oord2016wavenet}. 
Magenta, BachBot, and FlowMachines use input in the form of note sequences, while  Wavenet uses raw audio.


\section{User-centric and Collaborative Approaches for Artistic Creation}\label{sec:embodied}

The previous sections primarily demonstrate the role of computational arts for creative creation, solely driven by computers and deep learning. 
The achievements of deep learning among the novel artworks shed light on the paradigm shift. One possibility is that artificial intelligence achieves a high level of automation in art creations and eventually replace human artists. 
Nevertheless, we should not neglect the possibility of having human users who interplay with the computerised artistic agents. 
Within the creator community in the metaverse~\cite{Lee2021AllON}, it is very likely to see that human users with authoring systems work collaboratively with other AI-driven avatars or computer agents. Also, the metaverse can support enriched user interaction in immersive urban~\cite{csur-lee2022}. This envisions that multitudinous sensors capture users' physiological information and their movements. With appropriate privacy management of such data and informed user consent, such captured data can be further converted into artistic representations.

\begin{figure}[!h]
     \centering
     \begin{subfigure}[b]{0.3\textwidth}
         \centering
	\includegraphics[width=.8\linewidth]{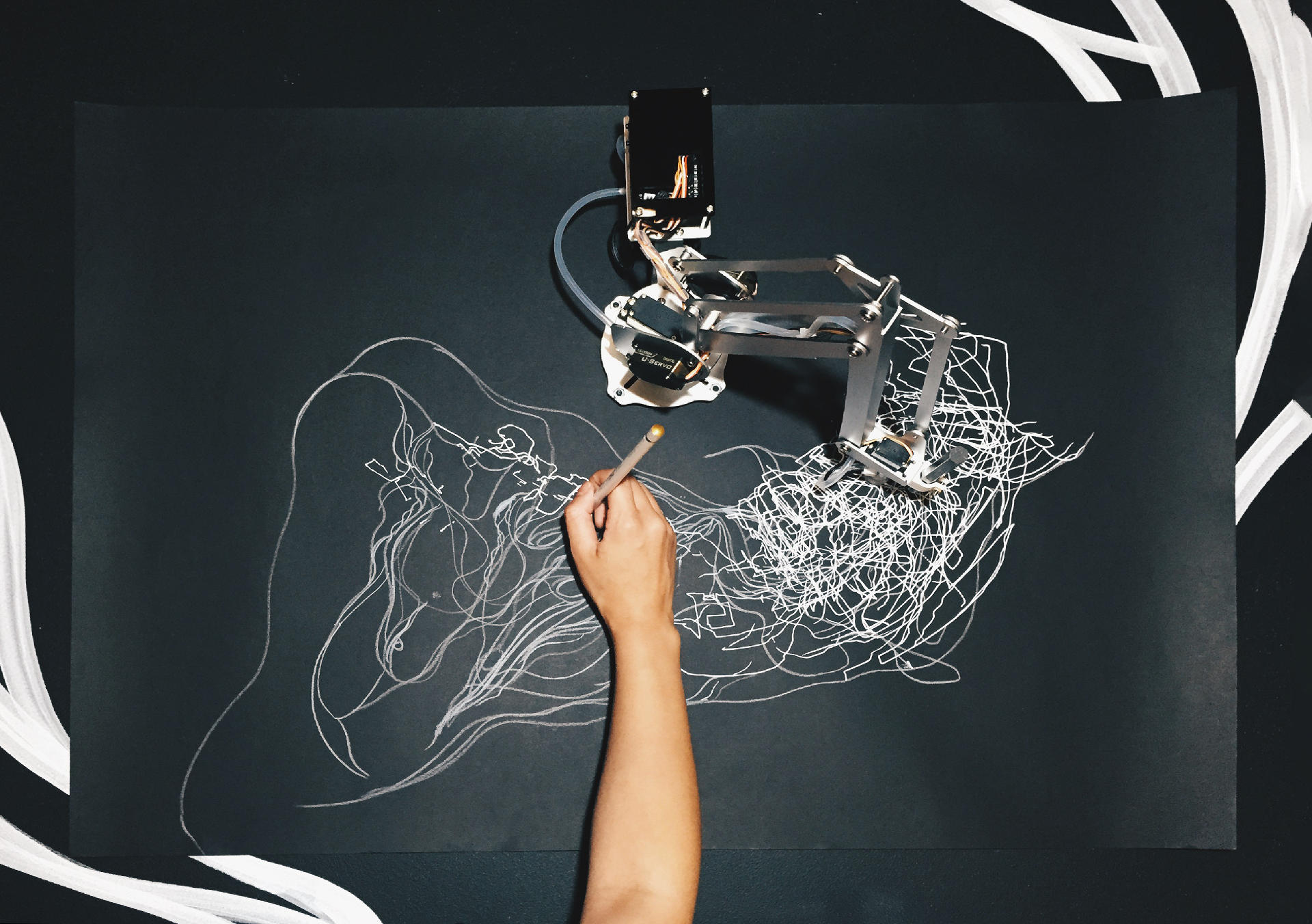}
	\caption{A human artist paints with a robotic arm.} 
	\label{fig:chung-robot}
     \end{subfigure}
     \hfill
     \begin{subfigure}[b]{0.3\textwidth}
        \centering
	\includegraphics[width=.91\linewidth]{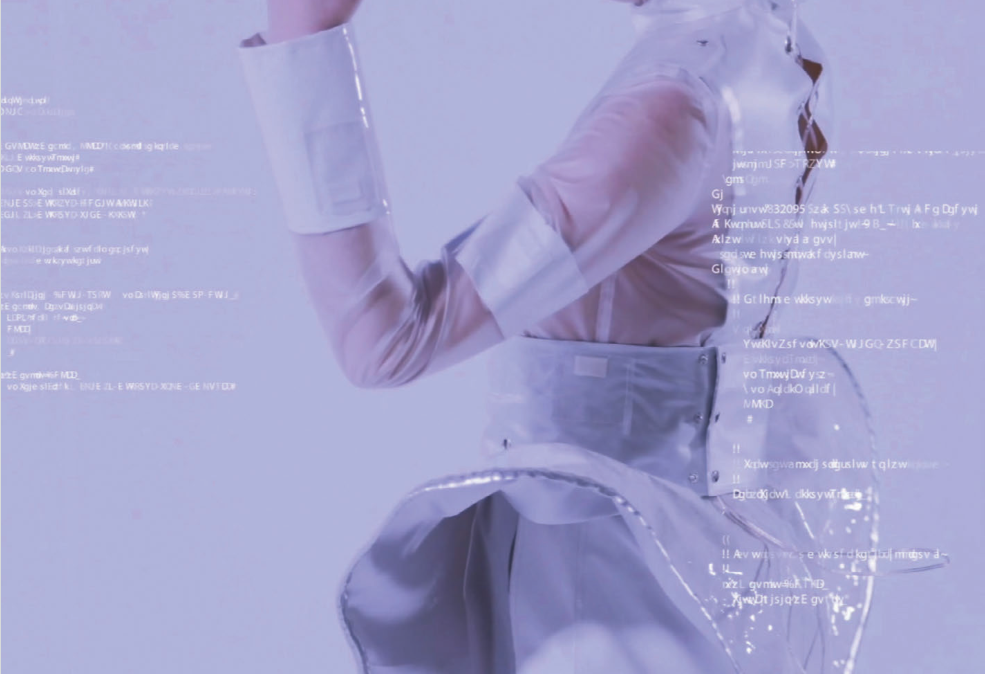}
	\caption{\textit{Body writing}.}
	\label{fig:Body writing}
     \end{subfigure}
     \hfill
     \begin{subfigure}[b]{0.3\textwidth}
         \centering
	\includegraphics[width=.9\linewidth]{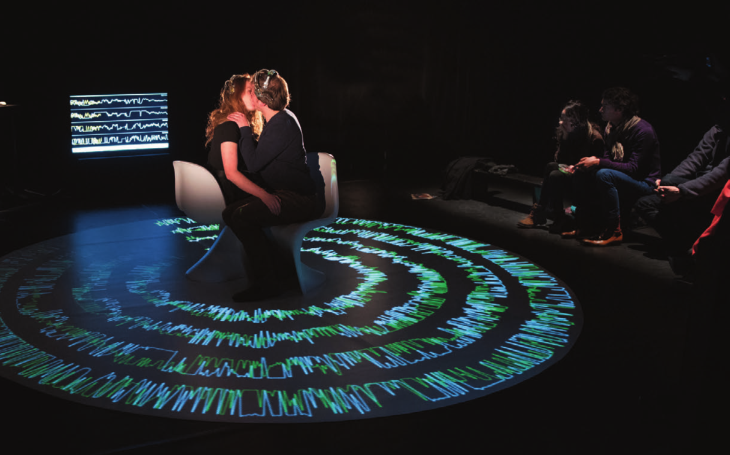}
	\caption{\textit{Sharing the Senses} by brain activities~\cite{Lancel2019EEGKS}.}
	\label{fig:Sharing the Senses}
     \end{subfigure}
        \caption{User-centric and Collaborative approaches for Computational Arts.}
        \label{fig:ssec8}
\end{figure}

The following paragraphs will discuss some examples that involve collaborative roles of computational arts and other user-centric approaches. 
As for the collaboration between human and computerised artists, AI-assisted collaborators, by leveraging robot arms and sensors to achieve physical embodiment, can work with human artists.
For instance, Artist Sougwen Chung\footnote{\url{https://www.washingtonpost.com/business/2020/11/05/ai-artificial-intelligence-art-sougwen-chung/}} is passionate about collaborative art-making with robots (Figure~\ref{fig:chung-robot}). Chung has designed and programmed about two dozen robots named Dougs, and they are guided by an AI system called recurrent neural networks. In order to make these Dougs her gestural expertise, she has uploaded her paintings online for over a decade. In addition, she has led AI-assisted painting performances on stages and in galleries before the pandemic. However, with the spread of the pandemic, Chung streams her robotic collaborations from her studio to replace live performances. 

As revealed by the above example, AI-assisted creation can help metaverse users to collaboratively complete certain artwork creation in virtual 3D worlds. Apart from finding AI-driven partners for creative processes, smart computation devices can surprisingly utilise the internal status of human users~\cite{Kwon2020MyoKeySE, Shatilov2021EmergingEN}, captured by body sensors, to create artworks. For instance, a wearable prototype named ``Body Writing'' (Figure \ref{fig:Body writing}), attached to 
a user's body and collects the user's biological statuses and emotions, connects the body with data through wearable devices. Based on the recognition of the user’s sadness, fear, enthusiasm, and other emotions, the body sensors inform AI artists, as user-centric inspiration, for writing poems. 

Furthermore, thanks to the advancement of machine learning that is capable of handling complicated signals, our brain activities being captured by Electroencephalography (EEG) electrodes, serving as brain-computer interfaces (BCI), can objectively mine, interpret, and classify social behaviours and emotion recognition. As recorded in a recent survey, numerous contemporary arts (N=61) have been completed by employing BCI from 1965 to 2018~\cite{Prpa2019BrainComputerII}.
Dutch artists Lancel and 
Maat initiated interdisciplinary research called ``\textit{EEG KISS}''~\cite{Lancel2019EEGKS} at the Art and Science Research Center of Tsinghua University in 2014\footnote{\url{https://zhuanlan.zhihu.com/p/94572785}}. With such sensing capability, we can leverage the ``\textit{Symphony of Intimacy Data}'' as artwork. 
As shown in Figure \ref{fig:Sharing the Senses}, the brain activity data of participants were captured by EEG smart wearables. When the two participants kissed and stroked each other's cheeks, the participants' emotions and the corresponding brain activity were 
converted into visualised data, as artistic representations, in real-time. 

\begin{figure}[!h]
     \centering
     \begin{subfigure}[b]{0.39\textwidth}
	\centering
	\includegraphics[width=0.98\linewidth]{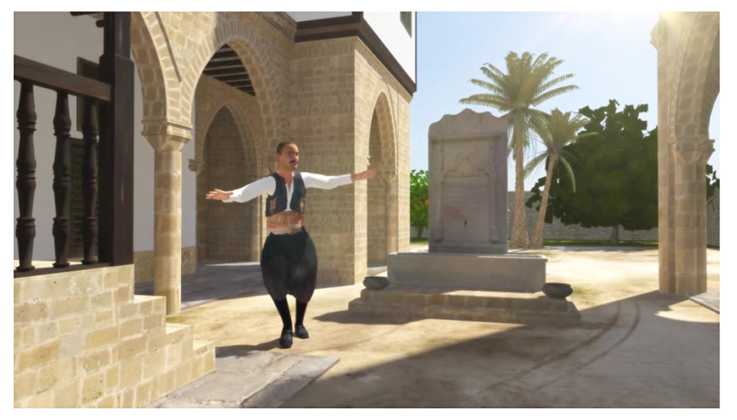}
	\caption{An animated virtual character with the Antikristos outfits and dance.} 
	\label{fig:An animated virtual character}
     \end{subfigure}
     \hfill
     \begin{subfigure}[b]{0.59\textwidth}
	\centering
	\includegraphics[width=.87\linewidth]{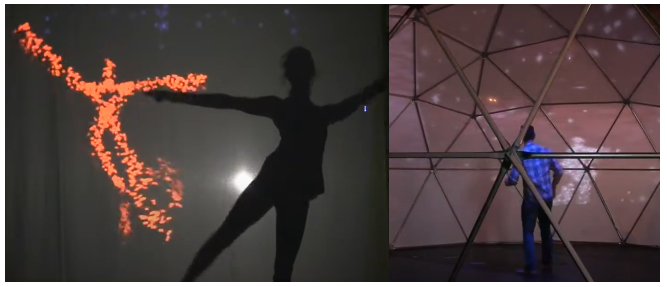}
	\caption{
	A participant (purple character) can explore the co-creative improvisation with the LuminAI agent (black character).} 
	\label{fig:An interactive tool}
     \end{subfigure}
        \caption{Virtual Avatars driven by the user's movements, and collaborative dances with an AI-driven avatar.}
        \label{fig:avatars-users}
\end{figure}

From the above cases, we can imagine that the internal status of metaverse users can bring some artistic effects to virtual 3D worlds. 
An imaginary example could be converting users' emotions or brain activities into the artistic aura to their virtual characters (i.e., their avatars). Another essential source of user behaviours is body movements and gestures, knowing that our body gestures could express emotions and moods, or even formulate a story through art performance like dancing. 
Back to the date before the prevalent use of machine learning, real-time computation of Fibonacci sequence and the golden ratio was regarded as the standard of judging the dancer's movement\cite{10.1145/1129006.1129018}. However, such quantitative metrics are hard to apply in enriched virtual environments. Lately, optical sensors (e.g., depth and RGB cameras), supported by convolutional neural networks (CNNs), are commonly employed to capture the dancers' poses and movements~\cite{8864522}. By employing common techniques like dynamic time wrapping (DTW)~\cite{10.1145/2617995.2618012}, the gestures and movements of the user (dancer) can be clustered as high-dimensional feature space and further converted into dance performance in virtual environments~\cite{10.1145/3344383}, as shown in Figure~\ref{fig:An animated virtual character}. 
In contrast, autonomous avatars can understand sound-motion mappings of human dancers, powered by deep learning, and hence mimic dancing movements with high levels of dance-likeness and emotional expressivity~\cite{10.1145/3450741.3465245}. 
Such sound-motion mappings, enabled by the Long short-term memory (LSTM) autoencoder, can be extended to the governance of artistic movements of virtual characters driven by certain musical rhythms in the metaverse~\cite{10.1145/3240508.3240526}. With Sparse-Temporal ReID Network, multiple dancers and their virtual postures can be re-constructed through image restorations~\cite{9054086}.



In the metaverse, computer agents, represented by autonomous avatars~\cite{Lee2021AllON}, can parse the user movements, and hence interplay with the human users' avatars. This can be further extended to the case of co-creation in the metaverse through dance performance. 
In a large display system of co-creative dancing named LuminAI \cite{10.1145/3401956.3404258}, 
human artists can perform dance improvisation (i.e., creative movement spontaneously). Then, LuminAI leverages unsupervised learning to learn about the user movements and formulate the human partner's \textit{gesture memory}. As a result, a LuminAI avatar appears to move and dance interactively (Figure \ref{fig:An interactive tool}). 
It is worthy of pinpointing that the dance performance is not limited to the dancers' body parts. Other factors such as space utilisation and dancer arrangements (i.e., their relative positions) should further be considered~\cite{10.1145/2000756.2000760}. 
Currently, dancing installation performance can be practically deployed at nightclubs and party gatherings. A prototype named `\textit{canvas dance}' can capture indoor activities of multiple dancers. The computers can detect multiple dancers' actions, and coordinate the dancers by giving visual cues in the form of blinking lights on a large display~\cite{10.1145/2702613.2732925}. 
Apart from dance improvisation, similar sensor and installation setups can support other types of interactive dance, such as Salsa (a Latin dance), Ballet, Contemporary Dance, and Waltz \cite{8468157, 10.1145/1565799.1565807, 10.1145/2702613.2725453}. 
Furthermore, the dancing art installations discussed above shed light on the possibility that avatars can understand the human users' movements, perhaps with a longitudinal observation in the metaverse, thus resulting in highly collaborative art performance across human users and autonomous avatars.
However, several unexplored areas exist in technology-inspired dance performance, including
connecting kinetics to the audiences, augmentation of expression for choreographers and dancers, aesthetic harmony of dancers as an aligned storyline, interactive build of the choreography, and integration between dancers (regardless of human or virtual characters) and interactive technologies/environments~\cite{10.1145/2399016.2399078}.

In November 2021, Roblox has released their ambitious investment plan valued 10 million US dollar for metaverse education. In the initial stage, three STEAM (Science, Technology, Engineering, Art and Mathematics) subjects, namely robotics, science, and space exploration, will be the first focal points\footnote{\url{https://www.wsj.com/articles/roblox-looks-to-bring-educational-videogames-to-schools-11636988400}}. The metaverse environments can further extend to dance performance to reinforce the Art. Regarding the feasibility, the latest technique of one-shot learning allows the conversion of human dancers into animated dancers with only one image. This opens the possibility for metaverse users from various backgrounds (e.g., body types, ethnicity, age, and gender) to quickly appear as virtual dancers~\cite{9578623}. Also, a prior work provides personalised learning and multimodal recordings (e.g., annotated videos and audios) to demonstrate the possibility of web-based platforms for remote interactions between dance learners, practitioners and experts \cite{10.1145/3212721.3212722}. In addition, the CHROMATA platform is designed to enable the scene understanding of cultural heritage and then build a digital twin of cultural heritage. Accordingly, such a metaverse scene can allow dancers to experience immersive dancing on the performance stage in a virtual 3D theatre of ancient Greece~\cite{9480948}.





\section{Physical Embodiment of Art: Robotics and Drones}\label{sec:robot}
In the era of intelligent industrialisation, we foresee the highly intelligent robots or robotic arms would achieve high levels of automation in production lines and hence superior production efficiency and quality. According to a report released by the World Robotics 2021 Industrial Robots report\footnote{\url{https://ifr.org/ifr-press-releases/news/robot-sales-rise-again}}, the number of robots increases by 10\% yearly, with the existing baseline of three million industrial robots or robotic arms currently operating in factories worldwide. Although Microsoft would like to pinpoint its metaverse with the workplace, the social robots, which may play a significant role in artistic creation and performance, are neglected. 
Although we mainly spot the usage of artistic robots appeared in the existing literature, human artists in physical environments can work collaboratively with artistic robots that either represent virtual avatars or AI-assisted agents from the metaverse~\cite{Lee2021AllON}. In addition, artistic robots present a new experience (known as a physical embodiment) to the artists during the creative process, in terms of environments, personal experience and emotions, and potentially inspire human artists to reach an unexplored landscape of art performance and artworks in the virtual-physical blended realities. 


In this section, we discuss diversified artistic robots designated for performance arts and artwork creation. First of all, musical robots can play traditional instruments or some adapted instruments. In addition to being able to perform the instructions given by the technicians, some robots have gained the ability to create new rhythms through machine learning algorithms. Noticeably, increasingly research effort is moving towards having robotic musicians play alongside human musicians. Therefore, in order to enrich the audience's experience of a live performance, the robot musician acquires additional interactive features from both audiences' reaction and performers. 
For instance, human musicians often move their bodies according 
to the tempo and intensity during 
the performance. Such captured motions allow the robots 
to be more in tune with the mood of human performers and audiences. 

\begin{figure}[!h]
     \centering
     \begin{subfigure}[b]{0.24\textwidth}
	\centering
    \includegraphics[width=.73\linewidth]{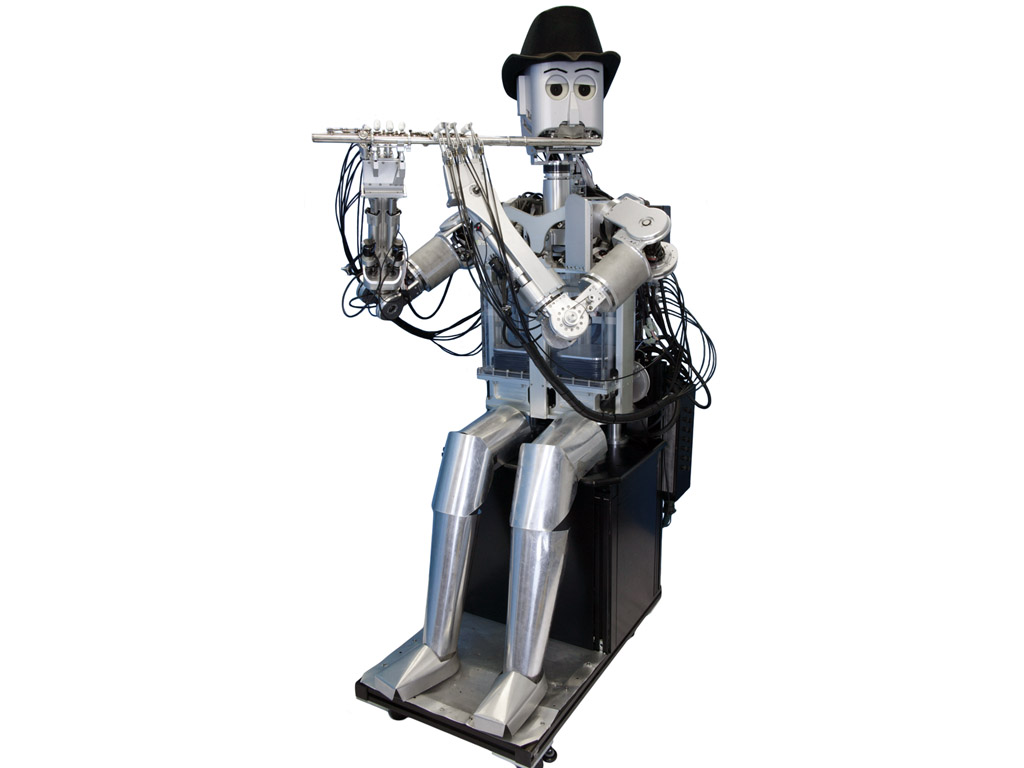}
    \caption{Waseda Flutist Robot.}
    \label{fig:flute}
     \end{subfigure}
     \hfill
      \begin{subfigure}[b]{0.23\textwidth}
	\centering
	\includegraphics[width=\linewidth]{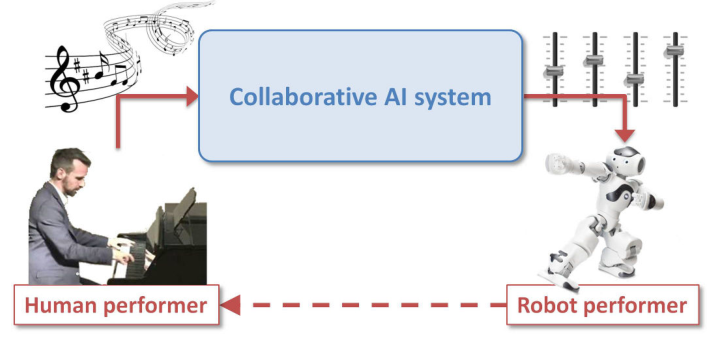}
	\caption{A pianist drives a robot's movements.} 
	\label{fig:A model for human-robot collaborative performance}
     \end{subfigure}
     \hfill
     \begin{subfigure}[b]{0.24\textwidth}
	    	\centering
	\includegraphics[width=\linewidth]{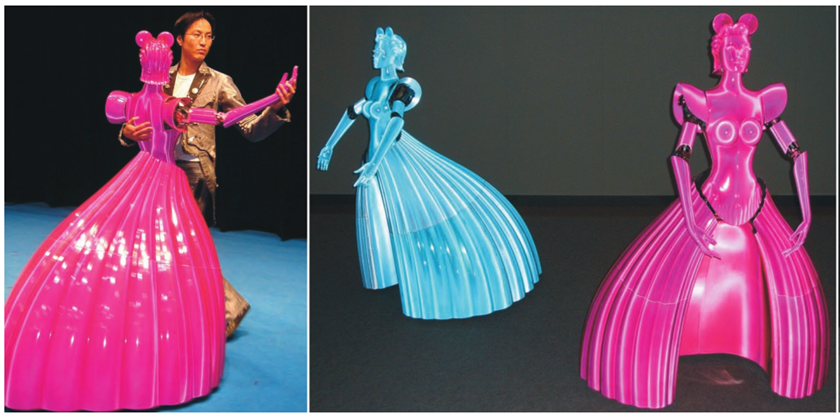}
	\caption{Dance Partner Robot}
	\label{fig:The Dance Partner Robot}
     \end{subfigure}
     \hfill
     \begin{subfigure}[b]{0.25\textwidth}
	    \centering
    \includegraphics[width=.6\linewidth]{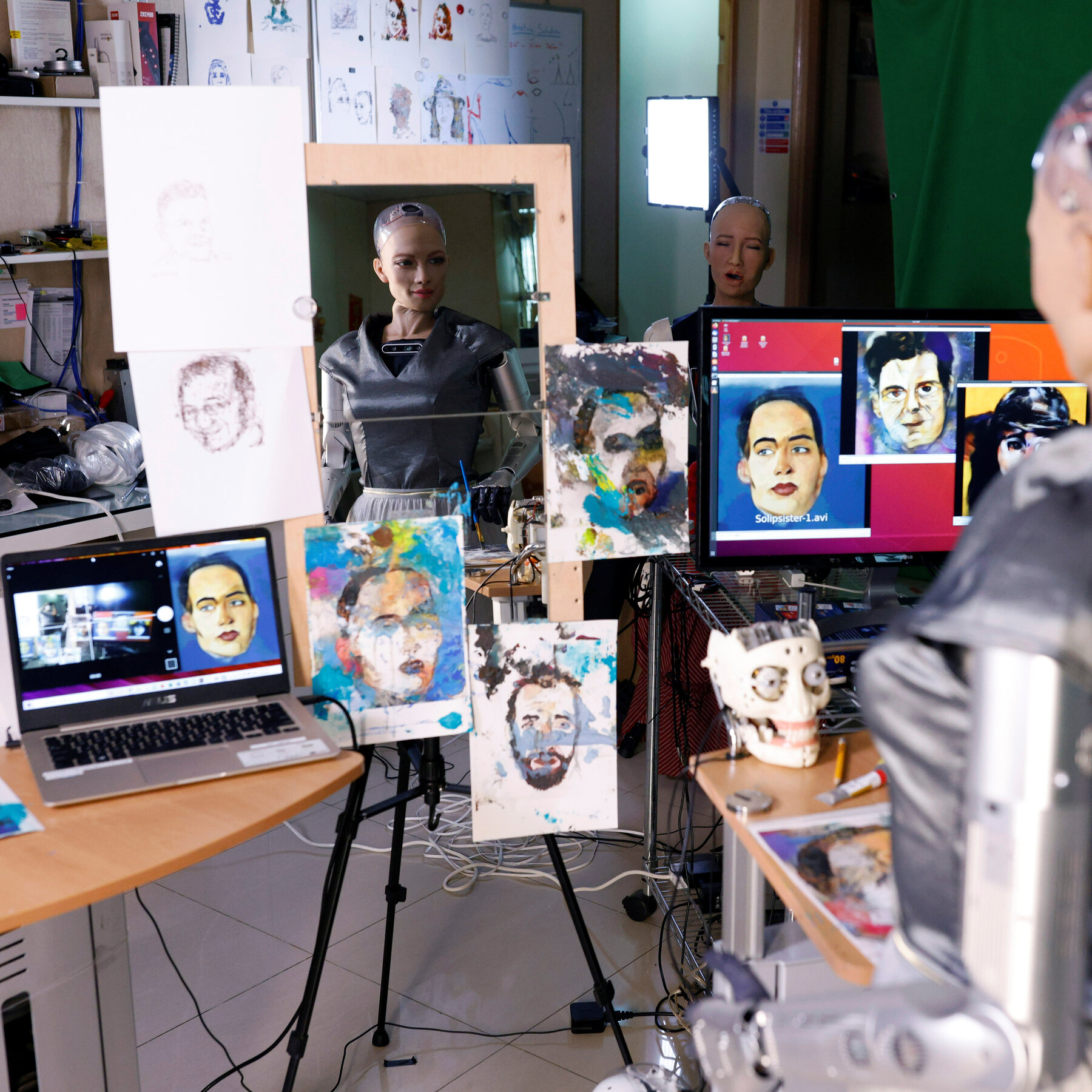}
    \caption{Sophia \& her paintings.} 
    \label{fig:sophia}
     \end{subfigure}
        \caption{Art performance by social robots, and collaborative performance.}
        \label{fig:social-robot-art}
\end{figure}

A robotic marimba player, namely Shimon \cite{hoffman2010shimon}, are designed to serve the aforementioned functions. 
Second, wind instruments, such as flute, are sounded by the player blowing into (or over) the instrument's mouthpiece, causing the air to vibrate. The shape of the mouth opening and the direction of the air blown in will influence the sound. The same principle applies to robot performers once the (soft) mechanics of the robots are available.
Figure \ref{fig:flute} shows an example of a flute robot. 
Musical performance robots are equipped with artificial lips \cite{solis2007musical} and air-supplying devices \cite{li2019designing} to improve their performance. Also, the robots can group and play music together. For instance, 
the conductor directs the musical performance in large group performances, such as orchestral or choral concerts. When musical robots join such events, they are able to recognise and respond to the conductor's gestures \cite{cosentino2012musical}.
Alternatively, fuzzy logic algorithms enable the robot performers to react to music performed by human, to achieve the human-robot joint improvisation performance~\cite{9223446} (Figure \ref{fig:A model for human-robot collaborative performance}). 

The integration of traditional dance art forms and robot technology gives dance art a new horizon and artistic meaning. With the increasing demands for robots to subtly convey emotions and gestures, robotics designers of dancing performance 
are pursuing the development of highly sensitive robots for complicated movements to enrich artistic performances. For example, enabling the robot to dance more naturally, such as changing its movement according to the user's real-time reaction, allows robots to express themselves through different dance movements. Therefore, more varied and characterful choreography becomes a crucial part. Dance robots can acquire new dance movements through machine learning or imitating movements from human dancers. For example, Nakaoka et al.\cite{nakaoka2005task} developed a bipedal robot that can imitate the movements of a human dancer of a Japanese folk dance. In addition, improvising dances allow dance robots to perform more human-like, including dance synchronised to music \cite{bi2018real}. Moreover, the development of robots with more flexible mechanical frames \cite{or2004biologically} could allow robots to perform more complex motions since there are no joints in the robot structure like in a human body. Dancers can express their feeling through their bodies to the audience during dance performances \cite{or2009towards}, giving robot dancers human-like characters.
In addition, artistic robots can dance interactively with other human artists~\cite{8864522} (Figure~\ref{fig:The Dance Partner Robot}). In particular, wearable technology like smart wristband that are equipped with IMU sensors (accelerator in particular) can detect the dancer motor movements and hence their skill levels of dancing movements~\cite{10.1145/3079628.3079673}. The captured dancer movements enable the robots to coordinate with the dancer's movements. 
With the deepening of human-computer interaction research, it is obvious that 
a variety of mobile robots can 
dance with humans ubiquitously. Robots with other physical form factors, like spherical robots \cite{10.1145/2875194.2875219}, can fit different needs of human performers. We would like to highlight that the performance with artistic robots is still in its nascent period. Performers can explore various robots and the artistic effects to create novel robot performances. 


As with other art forms, painters use graphics, composition and other aesthetic methods to express the concepts and meanings they wish to convey. There are many different types of painting, using media such as ink, acrylic and computer software. Robots can also draw in different styles \cite{luo2018robot,yao2005painting,gao2020making,seo2017picassnake}. Jean Tinguely presents his Meta-Matics machines, which can create unique abstract art by generating movement, at the Kaplan Gallery \footnote{\url{https://www.youtube.com/watch?v=VxoqVvQeil0}}. To the best of our knowledge, the painting created by the Meta-Matics machines could be considered as the very first robotic drawing. Apart from digital painting, most of the drawing process requires the use of brushes, paints and paper. The painter controls the strength and speed of the brushes to create different thicknesses of lines. Therefore, there are many studies on brush control approaches \cite{lam2011application}. For example, Yao et al. \cite{yao2005painting} studied the techniques of robot's brush control in Chinese painting, while Karimov et al. \cite{karimov2017brushstroke} presented a brushstroke rendering algorithm for ARTCYBE. Additionally, a painting robot, by Beltramello et al. \cite{beltramello2020artistic} presented a palette knife technique.

Furthermore, the shade of colour can indicate the contrast of light and dark and the distance relationship. Thus, the robotic painter also needs to consider these factors to make their drawings more human-like. In the case of coloured realistic paintings or black and white paintings, artists often mix pigments to create various colours to enrich the detail of the paint. As a result, robot designers created colour mixing devices \cite{karimov2019advanced,luo2016robot} for their robots.

Recent technology advancements have demonstrated that some drawing robots own a more human-like appearance and are able to communicate with humans -- e.g., robot artists Sophia \cite{hanson_robotics_2016} (Figure \ref{fig:sophia}) and Ai-Da \cite{meller}. Ai-Da is the world's first ultra-realistic artist created in February 2019. She draws with her robotic arm, AI algorithms and the camera in her eye. Moreover, Sophia is the world's first robot citizen, with her citizenship declaration 
in Saudi Arabia. 
In addition, she can communicate with people while painting, and recently sold her painting as NFT for \$688,888 \footnote{\url{https://www.nytimes.com/2021/03/25/arts/sophia-robot-nft.html}}.

\begin{figure}[!h]
     \centering
     \begin{subfigure}[b]{0.26\textwidth}
	    \centering
    \includegraphics[width=\linewidth]{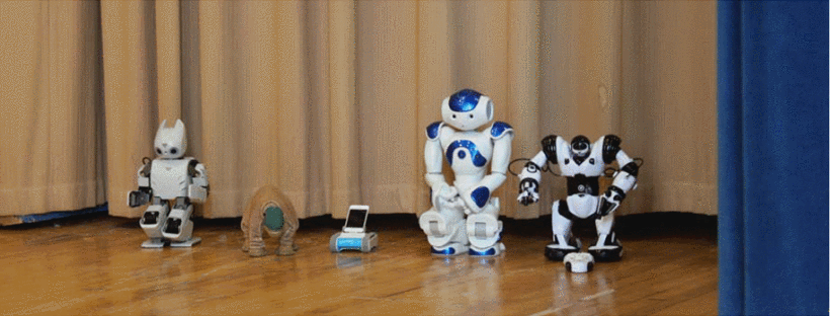}
    \caption{STEAM education with robots.}
    \label{fig:robotopera}
     \end{subfigure}
          \hfill
     \begin{subfigure}[b]{0.33\textwidth}
	\centering
	\includegraphics[width=.67\linewidth]{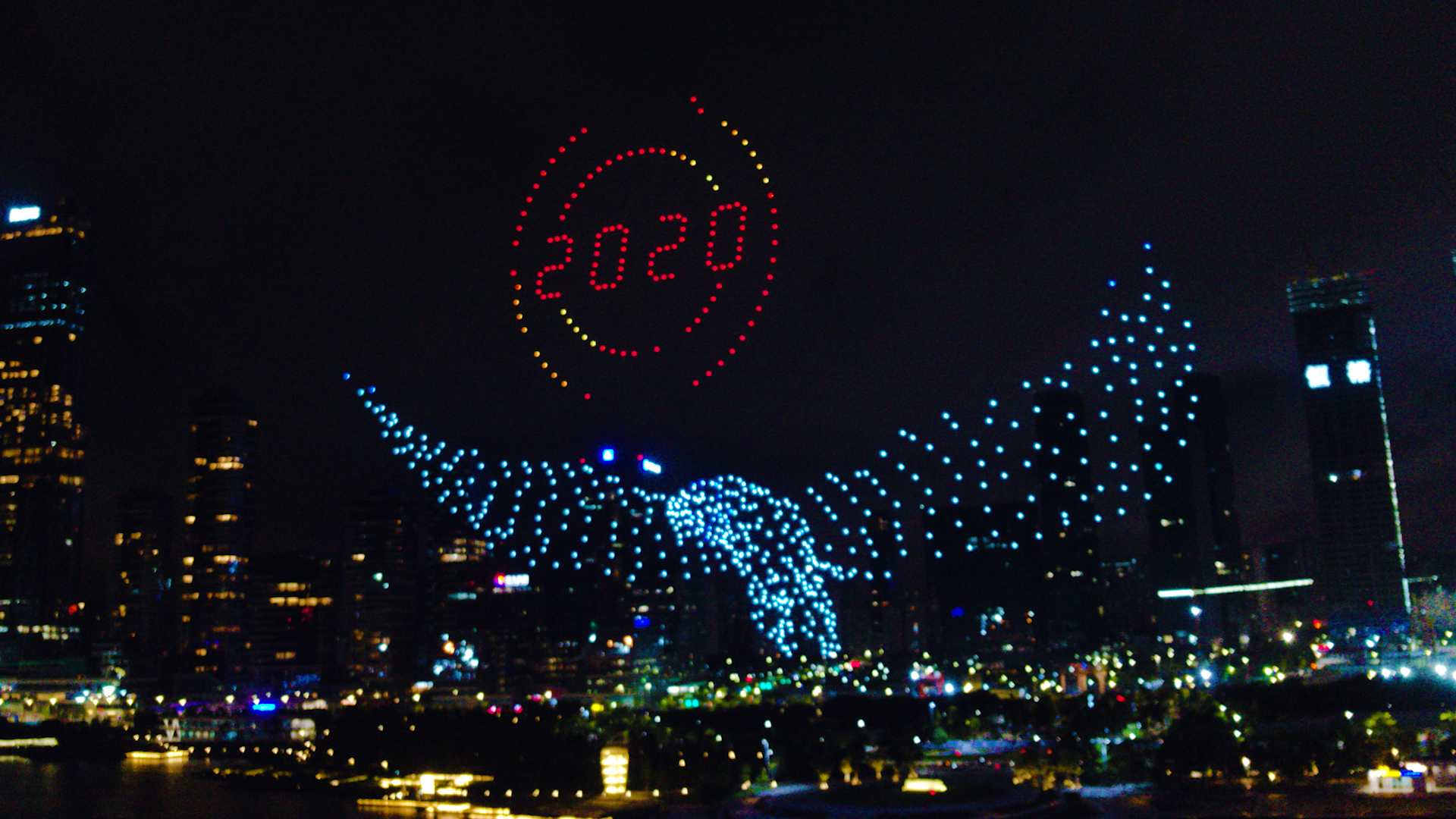}
    \caption{Light show by 826 drones\protect\footnotemark.}
    \label{fig:droneshow}
     \end{subfigure}
     \hfill
     \begin{subfigure}[b]{0.33\textwidth}
	\centering
	\includegraphics[width=.7\linewidth]{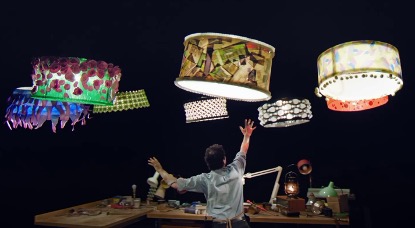}
    \caption{Live interaction with drones of costumes.}
    \label{fig:sparked}
     \end{subfigure}
        \caption{(a) Robotic theatre by social robots in an elementary school; (b--c) Flying drones' light shows \& arts .}
        \label{fig:fly-drone}
\end{figure}

A robot can also be a theatre actor. Theatre robotic actors not only brings a new form of theatre performance but also can use to inspire and teach STEAM (Science, Technology, Engineering, \textbf{Art}, and Mathematics) education \cite{jeon2017robot,ko2020robot}. Figure \ref{fig:robotopera} shows five robots used in a Robot Opera. Education in countries worldwide has now started to provide STEAM education to students at a younger age and cultivate the next generation of computational arts in the era of the metaverse. Therefore, many schools started to offer related subjects, such as programming and robotics, and some even started offering from elementary school \cite{jeon2017robot}. The advantages of using theatre robots for STEAM education are transforming unfamiliar content into familiar context and thus enhancing student engagement. Thus, getting more students interested in the art field through various STEAM activities. 
In addition to assisting education, theatre robots could stimulate more diverse research related to robots \cite{greer2019promoting,petrovic2019autonomous}, such as human-robot interaction (HRI). For example, helping to analyse positive HRI using the sensory system \cite{greer2017method}, and installing multiple robots into a facility using the Body-sharing multi-robot system \cite{umetsu2020body}.

\footnotetext{\url{https://www.youtube.com/watch?v=aIllsTyxt3Y}}

The use of drones for entertainment purposes could use for performance besides aerial photography. For example, drone shows \cite{youtube_2021} are one type of drone arts that uses multiple unmanned aerial vehicles (also known as drones) to fly in the air for display. The drone show is usually held at night, and drones are equipped with LEDs for various light effects; Figure \ref{fig:droneshow} shows a flying drone performance held in Shenzhen. Since drones are movable objects, they can be dressed up in costumes and performed with human performers~\footnote{\url{https://www.youtube.com/watch?v=6C8OJsHfmpI}} (see Figure \ref{fig:sparked}),  which provide a new form of performance for the audience. However, there are potential dangers in the use of drones if their design is defective. For instance, when the drone is flying, its propellers are operating at high speed, and it may cause injury if a person accidentally touches it during interaction with drones. Therefore, ensuring safety during indoor or outdoor drone shows is essential, and this requires a great number of technical engineering
\cite{kim2016realization,freistetter2020performance}. The drone will follow a pre-programmed route and use GPS or radio frequencies to determine its position and height with respect to other drones to prevent any collisions between drones. 

\section{Artistic Creations in Blended Environments and Virtual Creativity}\label{sec:ar-vr-art}

\begin{figure}[!h]
     \centering
     
     \begin{subfigure}[b]{0.17\textwidth}
         \centering
	\includegraphics[width=.7\linewidth]{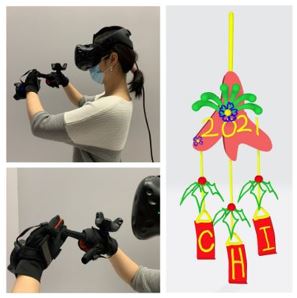}
	\caption{Hand drawing in VR.}
	\label{fig:vr-glove-lau}
     \end{subfigure}
      \hfill
      \begin{subfigure}[b]{0.17\textwidth}
     \centering
	\includegraphics[width=\linewidth]{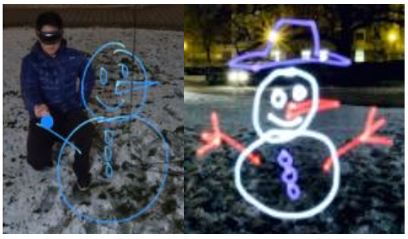}
	\caption{Using the lens of AR to create light painting.}
	\label{fig:light-paint}
     \end{subfigure}
     \hfill
     \begin{subfigure}[b]{0.17\textwidth}
         \centering
	\includegraphics[width=\linewidth]{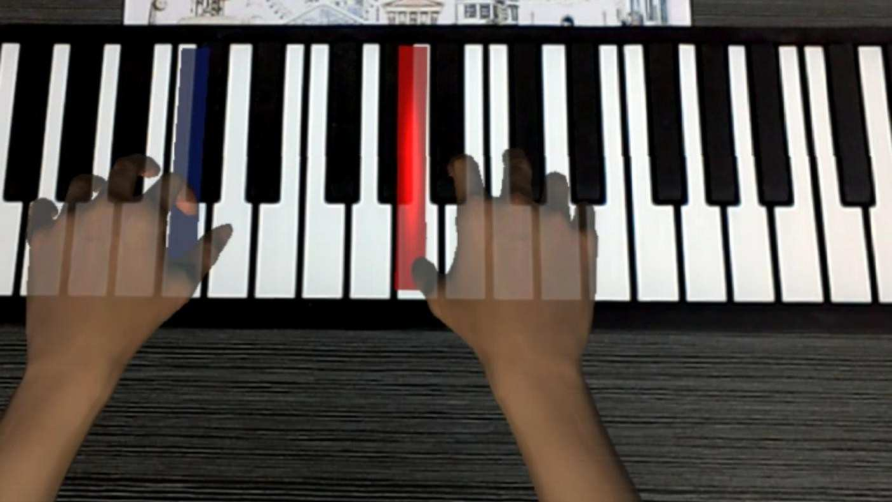}
	\caption{Playing the Piano with AR visual cues.}
	\label{fig:piano}
     \end{subfigure}
      \hfill
           \begin{subfigure}[b]{0.17\textwidth}
         \centering
	\includegraphics[width=\linewidth]{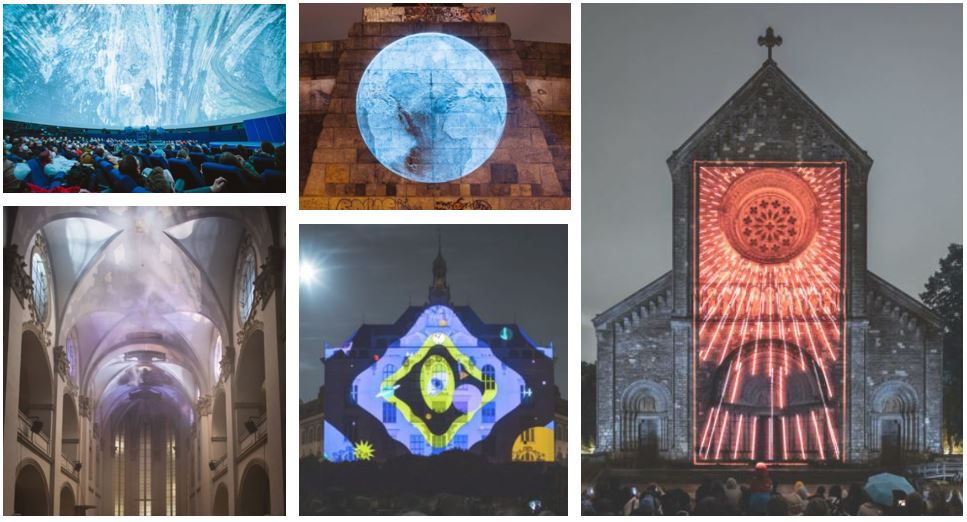}
	\caption{Playing the Piano with AR visual cues.}
	\label{fig:pan-m}
     \end{subfigure}
      \hfill
      \begin{subfigure}[b]{0.17\textwidth}
    \centering
	\includegraphics[width=\linewidth]{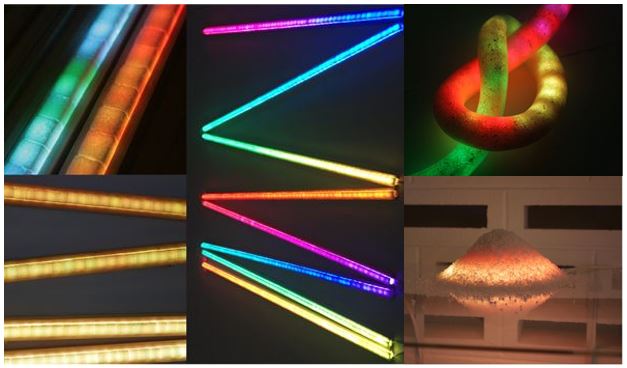}
	\caption{The weight of edible commodities.}
	\label{fig:data-c}
     \end{subfigure}
     
        \caption{Artworks in virtual-physical environments.}
        \label{fig:arvrarvr-art}
\end{figure}

The emphasised blending of virtual-physical realities for the metaverse cyberspace leverages virtual cyberspaces and physical spatial as a canvas for vastly diversified artworks. Examples include paintings and drawing in VR~\cite{lau-chi-2021-vrpaint}, creating AR overlays on top of physical environments~\cite{light-painting-AR} (Figure~\ref{fig:vr-glove-lau} and~\ref{fig:light-paint}), and decorating physical surface of heritage sites with AR and designing virtual buildings inside historic buildings~\cite{10.1145/3317552}.
Remarkably, \textit{Multiverse.pan}\footnote{\url{https://www.signalfestival.com/en/history/2019/multiverse-pan/}} utilises light projections to create immersive environments on buildings and their interior structures (Figure~\ref{fig:pan-m}). 
In addition, immersive environments can support and benefit art performances, e.g., offering visual cues when playing the piano~\cite{9479849} (Figure~\ref{fig:piano}). It is important to note that numerous immersive art installations consider arts and data together.
Though at first glance, data appears to be detached from the notion of artistic creation, the union of the two seemingly disconnected subjects has come into existence for a long time. Roughly 60 years ago, Jules Prown visualised his study in the history of American art with the use of punch cards~\cite{greenwald_2021}. However, data art as one of the fruitful products yielded from this union, extends beyond data visualisation. While the two strands both create visual entities that appeal to individuals' comprehension, they differ greatly in ends. 
Like the AR/VR projects mentioned above, immersive and light displays now serve as a conveyor of data arts that reshape our spatial environments. 
For instance, `\textit{Weight of Data}'~\cite{weight-of-data-chi} presents rice, sugar and salt as a light-display installation art showing raw crystal with a sense of data transformation (Figure~\ref{fig:data-c}). 

\begin{figure}[!h]
     \centering
     \begin{subfigure}[b]{0.36\textwidth}
	\centering
	\includegraphics[width=.79\linewidth]{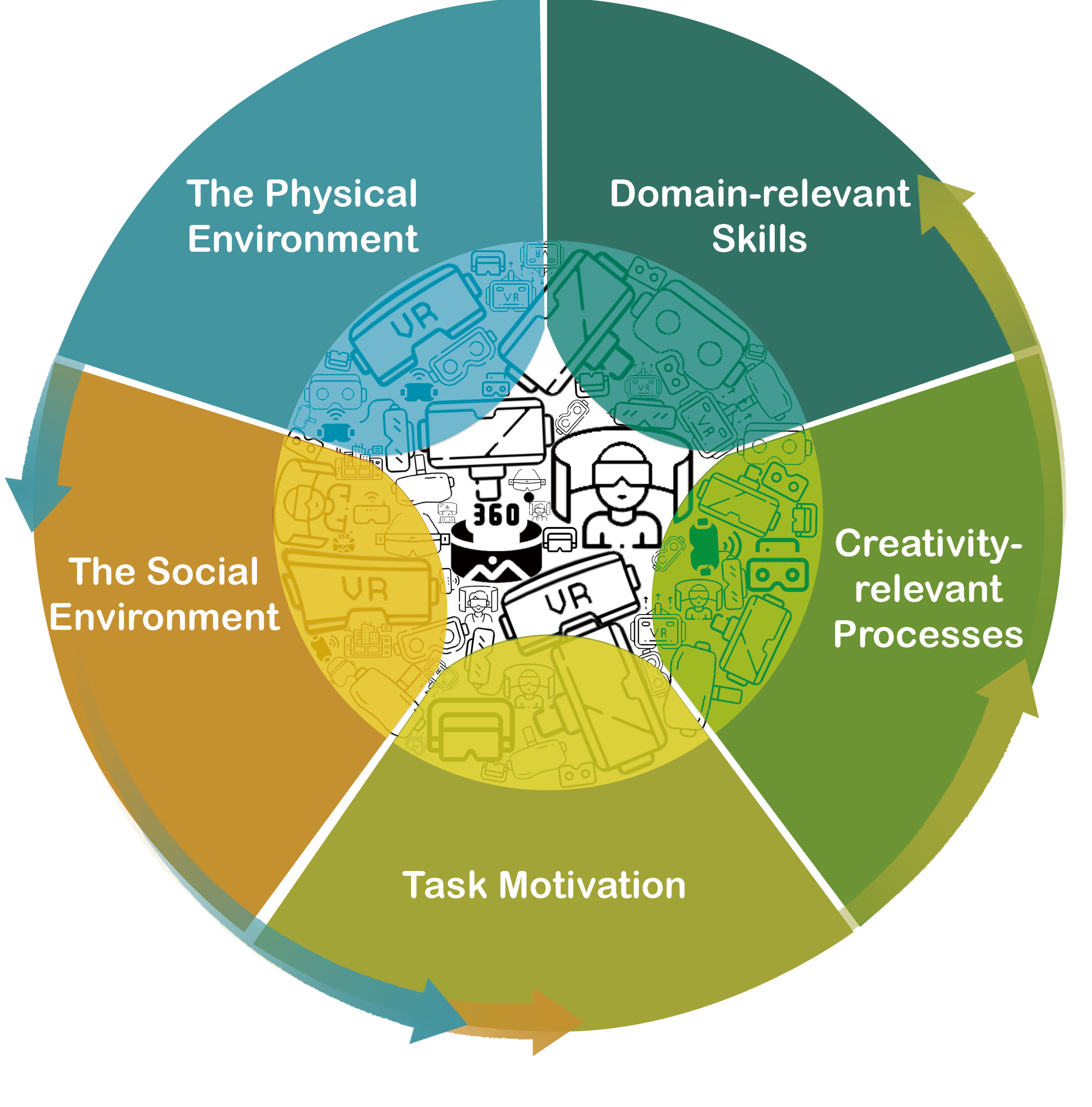}
	\caption{Interaction between VR and creativity components.}
	\label{fig:VR-creativity}
     \end{subfigure}
     \hfill
     \begin{subfigure}[b]{0.62\textwidth}
     \centering
	\includegraphics[width=\linewidth]{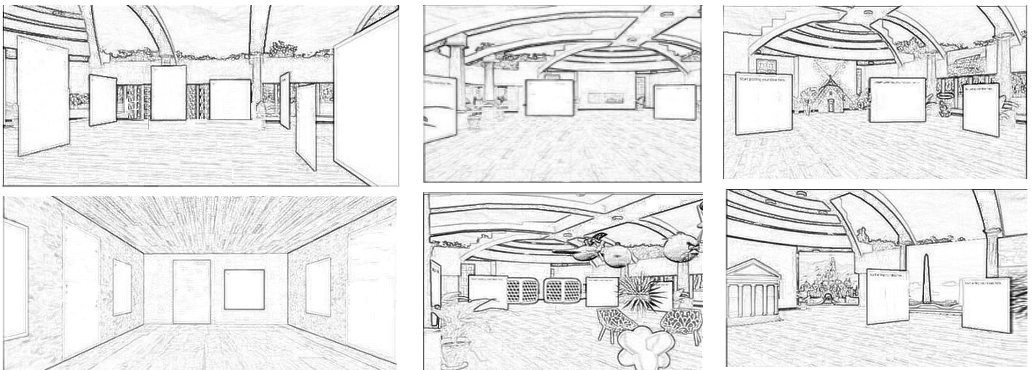}
	\caption{Three sketches of virtual creativity environments-- 1st column: the closing and opening environments. 2nd column: a virtual environment with/without specific 3D objects. 3rd column: topic-relevant virtual environment, source from \cite{minas2016opening, bhagwatwar2013creative, bhagwatwar2018contextual}.}
	\label{fig:ve}
     \end{subfigure}
        \caption{Creativity and Brainstorming in Virtual Environment.}
        \label{fig:virtual-creativity}
\end{figure}

As countless examples under artistic contexts exist, we cannot report all of them due to limited space. Instead, we focus on a prominent feature of nurturing creativity in virtual and immersive environments. 
Recent studies have investigated virtual worlds as a facilitating tool 
supporting individuals to improve creativity. 
By leveraging the virtual worlds, it is possible to nurture both the creative process and artwork creation in the metaverse. It is important to note that this concept aligns with the second stage as described in the three-stage metaverse development, namely Digital Native~\cite{Lee2021AllON}.
Due to creativity being a complicated conception, we explained the use of virtual environments such as AR, VR, and eventually the metaverse, in five aspects based on creativity referred to generate ideas or solutions that are novel and suitable for particular goals\cite{amabile2011componential}. According to the componential theory of creativity \cite{amabile2011componential,gong2020literature},  four components can affect creativity: domain-relevant skills, creativity-relevant processes, task motivation, and the social environment. In addition, although the componential theory of creativity is not including the physical environment, the previous studies verified that it could influence creativity\cite{thornhill2016virtual,gong2020literature,nilsson2006soundscape,minas2016opening,amabile2011componential}. Those five elements are influenced by one another, affecting creativity, as shown in Figure \ref{fig:VR-creativity}.


\footnotetext{\url{https://www.youtube.com/watch?v=aIllsTyxt3Y}}

Domain-related skills refer to the specific skills, knowledge, and capacity required to solve problems in a specific domain,\cite{amabile2011componential}, such as the drawing skills for artists. Integrating VR into practice could develop domain-related skills \cite{andone2018open}, such as the designing skill for designers~\cite{kim2020using,kandi2020application}. A study applied a VR application supporting students' garden designing and compared it to paper-based design \cite{kim2020using}. There are many advantages of using a VR application for garden design. For example, a specific factor in garden designing is the time dimension, which requires the designers to consider the influence of time, such as the daily changes by sun's position, seasonal changes, and the plants' changes by years. That is easy to be simulated in a VR application and provides the opportunities for designers to observe their design in different conditions and reduce the time needed in the real-world \cite{makransky2019motivational}, which helps designers practice and develop their designing skills in a virtual environment rather than the real world. Moreover, other skills also can be developed in a VR application, for example, the skills in design reviewing \cite{kandi2020application}. That is fundamental to identify design errors, such as missing components of products, 
and misuse of materials. Furthermore, for other domains, such as psychomotor skills \cite{alvarez2020use}, surgical training \cite{mao2021immersive} and nurse training \cite{chen2020effectiveness}, the researchers mentioned VR could simulate the real world to gain domain-relevant skills developing new ideas and solve problems, without incurring dangerous, expensive costs, and social problems. 

Creativity-relevant processes, or called creativity-relevant skills, consist of personality characteristics and cognitive styles, such as independence, risk-taking, openness, work style and skills, and the ability to break out of perceptual in generating ideas. VR is mainly applied in creative methods during creativity-relevant processes, which influence individuals generating responses \cite{amabile2011componential,gong2020literature,gong2021virtual}. For example, integrating VR into brainstorming and adopting the different avatars, improved the creative performance by measuring the fluency, originality, uniqueness, and novelty \cite{bonnardel2016enhancing,guegan2016avatar,guegan2017social,buisine2020proteus}. Specifically, the number of ideas (fluency) and the number of unique ideas (originality) could be increased by adopting an inventor avatar (such as a scientist) in a brainstorming task by applying VR \cite{guegan2016avatar}. In addition, the avatars with the identification (i.e., social identification) could contribute to the group creative performance and outcomes \cite{guegan2017social,buisine2020proteus}. Moreover, adopting VR in creativity-relevant processes could avoid shortcomings and have more advantages than without VR \cite{petrykowski2018digital,kutak2019interactive}. The shortcoming, such as saving the conversation or capturing the discussion contents, is easily solved by adopting a VR application \cite{morales2020syncmeet}. 
For example, it is easy to undo an operation in VR application allowing the participants to practice trial and error in design processes not feasible in the real world, which supports creativity-relevant processes to improve creativity \cite{kim2020using}.   

Based on the componential theory of creativity, task motivation specifically refers to intrinsic task motivation that is the degree of personal enthusiasm for completing tasks or solving problems. For example, challenging or exciting tasks can stimulate self-challenge and interest and satisfy the individual \cite{amabile2011componential}. Some researchers have suggested that the immersive environment created by VR could be a valuable way to increase personal interest \cite{kivela2019study, ijaz2020player}, and learning motivation \cite{lin2021using,al2020effectiveness,bernardoni2019virtual}. The researchers mentioned that immersive systems, including the metaverse, are interesting and inspire motivation and creativity \cite{lee2019can}. Another study also emphasised that individuals worked hard to create a solution in an immersive environment, and they found it more fascinating, and dramatic than an ordinary physical environment, \cite{bujdoso2017developing}. In addition, users are satisfied with the process of creation, and they believe that the virtual environment was accompanied by dramatic challenges that motivated them to perform better, thereby enhancing creativity \cite{han2019virtual}. Moreover, the VR game, for instance, through priming game \cite{dennis2013sparking}, can motivate an individual by delivering illustrations of the goals and achievements to spark the individual’s creative performance and eventually gain a big pool of ideas. Furthermore, a study verified that playing VR games encourages individuals to engage in training and motivates them to exercise more often \cite{kivela2019study}. The studies verified that applying VR to complete a task or solve a problem or learning or training could stimulate individuals' motivations, capture their interests and make them engage in the processes.

The social environment, perhaps socialised virtual worlds, generally includes all extrinsic motivators, which might hinder or stimulate intrinsic motivation and creativity \cite{amabile2011componential}. Some extrinsic motivators might be obstacles to creativity (e.g., excessive time pressure and demanding leaders), which is in stark contrast to a friendly social environment where free sharing of comments and similar motivations can stimulate creativity. For example, to participate in a task anonymously in a virtual environment, which supports an equal social environment. Participants can freely express their ideas and provide feedback, which encourages them more voluntarily to share their ideas, and more free communication and avoids several negative external motivations such as a strict hierarchy, thus improving creativity \cite{fominykh2012supporting,ide2020effects,kutak2019interactive}. Other extrinsic motivators, such as collaborative exploration, co-modification, and mutual co-exploration, encouraged individuals to generate original solutions and inspired them to discover unexpected new solutions \cite{hong2016enablers}.  Moreover, VR could also be used to create a virtual environment (e.g., situated learning environment), contributing to enhancing learners' motivation \cite{mei2011applying,huang2020learning}. The participants expressed their feelings and provided their feedback in the interview after using a spherical video-based virtual reality supporting teaching, such as the presence and immersion, which encourage them to be more focused in learning and stimulate their learning motivation (i.e., extrinsic motivation) \cite{huang2020learning}. 

The physical environment refers to the geographic area and factors around human beings \cite{moo2019comprehensive}. It is worthwhile to mention that the metaverse will contain a huge number of creators or artists across different cultures and geographic areas. Some specific factors in physical environments to be verified could enhance creativity \cite{thornhill2016virtual,dul2011work,leung2012embodied}. Therefore, the researchers simulated the physical environment and added specific factors by the use of augmenting such environments with virtual entities (i.e., mixed reality), 
evaluating the simulated environment that could be used to inspire creativity. Comparing the closing environment (without windows and closed roof), the opening environment with arched glass roofs (left column, Figure \ref{fig:ve}) enabled individuals to contribute more original, and practical ideas to teams \cite{minas2016opening}. Also, the virtual environment with specific 3D objects (e.g., specific trees and flowers), could influence individual cognition, thereby enhancing individuals’ creativity \cite{bhagwatwar2013creative} (the lower illustration in the middle column, Figure \ref{fig:ve}). Moreover, the prior study 
continued to investigate this topic by comparing the topic-relevant virtual environment (right column, Figure\ref{fig:ve}) and the specific virtual environment mentioned above. Their results showed that both virtual environments could improve creative performance because priming can manipulate individuals' unconscious cognition and behaviour to improve team performance. In particular, the topic relevant environment is a suitable method for implementing priming in VE, which can encourage personal design thinking, especially in 
ideation stages 
\cite{bhagwatwar2018contextual}.


Although the researchers put lots of effort into VR to enhance creativity, there is much room for further exploration. For example, lacking facial expression is still a neglected factor that negatively affects users' experience in the creativity-relevant processes. Adding other senses also a promising method, such as the smell of fresh air, enabled individuals to involve in the creative process peacefully and calmly, which influenced their cognitive, and positive emotions related to task motivation and the social environment enhancing creativity \cite{plambech2015impact,gong2020literature}. The advent of the metaverse, perhaps with more intelligent and comprehensive sensing capability, can further open the opportunity to enrich the environments of virtual creativity.

\section{Research Agenda}\label{sec:discussion}

The article surveys numerous recent examples of computational arts with obvious concatenation with the metaverse. 
We project that innumerable artworks will appear in the metaverse era, considering that our everyday objects in the physical reality would turn into an endless canvas. 
As such, artists will leverage various opportunities from novel computational devices, new materials and media, and mixing of virtual entities and physical environments, to express their idea, messages, and creativity, being conveyed by artworks. 
Instead of deepening the discussion of various artwork and creative processes, the grand challenge of metaverse creation primarily links to the design of the metaverse ecosystem and its governance. Some fundamental questions could be: how the metaverse environments encourage content creation by every metaverse user? Also, how to preserve the metaverse creation (e.g., digital heritage)? And, how to achieve a creator economy,  like nowadays virtual art trading? In addition, how to ensure the privacy and safety of metaverse users, especially those who are vulnerable like underage content creators, and help artists understand the social impacts of their works?
We consider such ecosystem issues are critical to 
accelerating the momentum of metaverse creation. Thus, we outline a research agenda to reflect several urgent matters regarding establishing a metaverse ecosystem for art and culture activities.


\paragraph{Democratising Computational Arts and Education} The year 2021 marks the twentieth anniversary of the Processing project, an open-source, Java-based programming framework initiated by Casey Reas and Ben Fry \cite{processing-website}. It is designed with the goal to enable artists and designers to more easily employ computer programming for visual media and creative output \cite{10.1007/s00146-006-0050-9}. Over the years, Processing and other similar creative programming frameworks become vastly popular, with their user bases no longer limited to art and design practitioners. Not only have these frameworks made programming accessible to a wide range of fresh learners, those who have formal training in computer programming are also starting to experiment with these frameworks to create visual outputs and thereby engaging with art and design directly. This close interaction between the arts and computation gave rise to a new community that blends people from both backgrounds. These phenomena reflect an important trend in recent years, which is the democratisation of computational arts and education. Creative programming frameworks, with their open-source nature and ease of use, have made it possible for most people to start creating multimedia output in a relatively short time. This trend is not only seen in programming, but also in the field of CGI and games, exemplified by the open-source 3D content creation software Blender \cite{blender-website} and the professional-grade Unreal Engine that is mostly free and also has its source code openly published \cite{unreal-source}. If this perceived trend continues, it is expected that they will bring even greater energy into the computational arts field, though the details of their impacts remained to be further studied.

\paragraph{Barriers of Building Virtual Scenes and Characters}
In the field of CGI and video game, independent creators now have access to relatively easy-to-use and inexpensive or even free software, and are able to be self-taught through a large amount of tutorial resources on the internet. Many of the assistive processes and tools, such as motion capture, are also becoming more accessible, so that what used to require a team of dozens of people can now be done by a minimal group or even an individual, greatly stimulating the enthusiasm for creativity. At the same time, there are still issues that need to be addressed and further researched. First, the degree of integration between process and platform still needs to be improved, as content creation platforms often have their own focus areas, and therefore users often need to transfer and process content between multiple platforms in the metaverse, but this process considering the interoperability is sometimes quite cumbersome, not to mention the need for possible future integration to more interactive and networked presentation platforms. Second, software usability still can be improved, especially as most of the software is still designed for team-based, professional pipeline, and although it has become more user-friendly for individual users than before, some still has a steep learning curve, as we expect that every participant in the metaverse can enjoy the democratisation of content creation.

\paragraph{Digital Privacy and Safety for Artists in the Metaverse. }In online communities like Transformative fandom, content designers share their works that have been derived from the original content. For example, members of the \emph{Harry Potter} fandom can write their own stories on the on-screen relationship between Harry Potter and Hermione Granger. Since the work is creative, it often goes beyond the social norms which are generally accepted in real life. So, it becomes imperative to ensure privacy and safety for content creators. A Recent work by Dym et al. \cite{dym2020social} reports that members of such communities often rely on implicit understanding from the fellow members for ensuring privacy and safety, and often such threat arises from new members who have recently joined the community, but do not conform to the implicit understanding. Given the recent report on harassment and cyberbullying in VR social network platforms \cite{vrabuse2016}, new mechanisms will need to be developed which can tackle these issues, and especially protect the most vulnerable members, like underage artists. Another concern arises from the perspective of digital content moderation. 
Since some artists are uniquely positioned to challenge norms of society, whether it is social, political, or economic, it is important to equip them with tools that can help them to understand the potential influence of their contents on society at large. The most common strategies adopted by the society or the government to control such influence are either community moderation, or censorship. However, norms based control measures do not have rooms for including the ideas which are not widely accepted by the society, and which artists often tend to bring-forth through their arts.

\paragraph{Digital Art Trade Beyond Ownership Recognition}
The emergence of NFTs kindled the vision of a new business model for digital arts. The capability to establish ownership, however, is not a sufficient condition to produce an effective framework that successfully imitates how traditional art markets works. One area worth addressing is leasing NFTs for personal and commercial purposes. In traditional art markets, a piece of artwork curated by an individual collector or an art gallery may be leased by others for time-limited display or other commercial events. By similar analogy, metaverse residents who hold precious virtual artworks may also face scenarios where they receive requests for leasing their collections. Currently, one notable step taken forward in addressing this issue is the reNFT project\footnote{\url{https://www.renft.io/\#about}}, which proposes a rental protocol that enables NFTs owners to profit from leasing their collections to others. Nonetheless, one should recognise the need for collective efforts in enforcing a functioning digital rental market in the metaverse. Specifically, one needs to scrutinise key questions, including how disputes arising from the lease of NFTs should be settled, which may require institutional inputs from metaverse creators. In addition, a similar complexity arises from the different types of rights involved in transactions of the artworks from artists to buyers. The current technologies allow artists to specify what rights of a particular piece of artwork (e.g., the right to commercial use) are to be transferred when minting the token. Nevertheless, it is still important to ensure both artists and buyers understand the legal implications of the terms and act accordingly. Once again, the need for sufficient enforcement for terms specified in the NFT licenses to be honoured immediately becomes a salient issue. Another challenge arises from the fact that all digital goods (and hence digital arts) can be copied, which is detrimental for content designers in the metaverse. Even though a potential customer would be legally bound to the terms and conditions agreed with the content creators before leasing, we need to ensure that the customers cannot make a copy of the art on the technological level. Time-limited attestation of the digital art with the verified VR hardware of the customers could be one potential solution to address this concern. 
Also, limited research has been done on environmentally friendly NFTs. As it is undesirable to create a virtual world built upon technologies that backfire on our physical home, we encourage readers to consider how NFTs can be used in a sustainable manner.

\paragraph{
Intangibility of Metaverse Creation.} The current music technologies for the metaverse have many limitations. For example, artists and musicians are used to feeling physical pressure on their fingers when they play instruments in the real-life, and this pressure channels the music sensations throughout the body. Current VRMIs do not have such feedback mechanisms. One of the possible solutions could be to utilise a brain-computer interface that can channel the sensations to musicians. Another approach would be using Haptic Gloves, like the ones proposed by the Facebook Reality Lab \cite{fbglove}. Music concerts in the metaverse though offer an alternative to physical concerts and an opportunity to bring multiverse themes at the concert stage and hall, it needs to tackle a few issues, i.e. audience's avatars may wander to restricted space, the projection of the audience's avatar may overshadow the musician's avatar, as observed in the performance of Jean-Michel Jarre on VRChat platform in June 2020 \cite{jarre2020}. As for supporting artists in creating (creative) music, deep learning-based approaches have mostly seen success in classic genres. Other genres are ripe for research. All in all, artists who have given performance in the virtual world are still not using VRMI, for instance, musical metacreation (Section~\ref{sec:auditory}), but physical instruments, even though VRMI could provide a richer choice to compose music in an immersive environment, which opens a door to develop training programs that can help artists trained with physical music instruments, adapt to virtual reality music instruments.

\paragraph{Turing Test for Computational Arts.} Turing test provides a generic framework to measure the machine intelligence by asking a simple question: \emph{Can machines think?} Extending this idea to the domain of computational arts, the question becomes \emph{Can machine think as well as be creative?} Having a generic framework to answer this question is imperative, since we are likely to see novel computational arts in the future cyberspace of the metaverse which we would need to assess.
The widely accepted answer considers two factors: 1) An unbiased human can not distinguish whether the given artwork (generated with aid of the machine, but this fact unknown to the judge) has been generated by the machine, or by the human, 2) The given machine-generated artwork has as much aesthetic value as one produced by a human artist. If both of these factors are affirmative, then such an artwork generated with machine-aid could be considered creative. However, this framework has several loopholes. Should the unbiased human judge be an ordinary Joe/Jane or an expert in the field? It has been reported that on several occasions the audience withdrew their previous valuation of the image and music after coming to know that they were computer-generated \cite{boden2010turing}. The crux lies in the perception that computer-generated art is not really art, and any beauty it may have is purely superficial. This perception arises because, on the most fundamental level, the computer has to follow human instructions to generate anything which diminishes the essence of "beauty" and "novelty". To tackle this issue, we would need to think to the next level: \emph{Can computers create?} There is a subtle but significant difference between this question and the previously mentioned question of being creative because the baselines for answers to each of those two questions are distinctive. In the former question, the baseline of an unbiased human judge not being able to distinguish may suffice, however for the latter question, it wanders in the domain of computer being able to create something new without any human (or artist) instructions and we still need to come to an answer in the metaverse era. 

\paragraph{Technological Infrastructure of Metaverse Creation.} The maturity of input and output devices for AR/VR, mobile network, and edge and cloud infrastructures would facilitate the 
metaverse creation at scale~\cite{Lee2021AllON, Zhang2021EdgeXARA6}.
Generative art and other automatic generation methods of artworks involve demanding computational tasks. Due to insufficient computation resource and short battery life~\cite{Lee2020UbiPointTN}, wearable and mobile computers, such as AR and VR headsets, strongly rely on task offloading~\cite{Zhang2021EdgeXARA6}. The real-time generations of calligraphy and poems, especially a great number of creators that exist in the metaverse~\cite{Lee2021AllON}, would be challenging. 
Furthermore, the latest development of input devices for user interactivity in virtual-physical blended worlds are tackling the bottlenecks of throughput rates~\cite{csur-lee2022} and investigating alternative modals for delivering sensation for user-virtuality interaction~\cite{carlos-hexa}. 
Creating artworks in such environments would require more advanced devices to convert user inputs into the desired expression, e.g., the acceleration of the user's arms can reflect a stroke width in mid-air drawing with a virtual brush. We expect technologists or interaction designers to employ varied sensors and multitudinous form factors to pursue subtle and accurate devices or tools for metaverse creation. 
Also, the limited field-of-view (FOV) on the display screen of the latest AR and VR headsets~\cite{seen-to-unseen} could impact the user immersiveness 
and hence potentially the appreciation of visual arts. 
Such FOV constraints encourage designing immersive recommendation systems that offer precise yet highly personalised artistic content to the audiences~\cite{Lam-A2W-MM-2021}.
 



\section{Concluding Remark}
With the burgeoning virtual art trading, the computational arts's next decade will look radically different from what it is today, driven by the advent of the metaverse. Metaverse cyberspace will open numerous opportunities for creators and artists to reshape our virtual and physical environments in artistic and novel ways. 
By surveying the most recent works across virtual photography, cinematic simulation, calligraphy, poetry, musical metacreation, immersive arts and virtual creativity, as well as other artworks driven by user embodiment and robotics, we hope to have offered a broader discussion within the community of computational arts. 
We pinpointed the above vital topics and discussed the research agenda and several fundamental challenges to construct an artistic vision for the metaverse cyberspace.
We call for interdisciplinary research requiring significant efforts from both technologists and artists to co-investigate the integrated facets of computational arts and technological infrastructures of such an artistic metaverse.


\bibliographystyle{ACM-Reference-Format}
\bibliography{main}

\appendix









\end{document}